\documentclass[lettersize,journal]{IEEEtran}
\usepackage{amsmath,amsfonts}
\usepackage{algorithmic}
\usepackage{array}
\usepackage{textcomp}
\usepackage{stfloats}
\usepackage{url}
\usepackage{verbatim}
\usepackage{graphicx}
\usepackage{subcaption}
\usepackage{subfig}
\usepackage{multirow}
\usepackage{caption}
\usepackage{booktabs} 
\usepackage{float}
\def\BibTeX{{\rm B\kern-.05em{\sc i\kern-.025em b}\kern-.08em
    T\kern-.1667em\lower.7ex\hbox{E}\kern-.125emX}}
\usepackage{balance}
\usepackage{enumitem}
\usepackage{graphicx}
\usepackage{titlesec}
\titleclass{\subsubsubsection}{straight}[\subsection]
\titleformat{\subsubsubsection}[runin]
{\normalfont\normalsize\bfseries}{}{0em}{} 
\titlespacing*{\subsubsubsection}{0pt}{3.25ex plus 1ex minus .2ex}{1.5ex plus .2ex}
\makeatletter
\def\toclevel@subsubsubsection{4}
\def\l@subsubsubsection{\@dottedtocline{4}{7em}{4em}}
\makeatother
\newcounter{subsubsubsection}[subsection]

\begin{document}
\title{Deep Learning Based Forecasting-Aided State Estimation in Active Distribution Networks}

\author{
    \IEEEauthorblockN{
        Malek Alduhaymi, Student Member, IEEE; 
        Ravindra Singh, Senior Member, IEEE; 
        Firdous Ul Nazir, Senior Member, IEEE; 
        and Bikash C. Pal, Fellow, IEEE
    }
}



\maketitle

\begin{abstract}
Operating an active distribution network (ADN) in the absence of enough measurements, the presence of distributed energy resources, and poor knowledge of responsive demand behaviour is a huge challenge. This paper introduces systematic modelling of demand response behaviour which is then included in Forecasting Aided State Estimation (FASE) for better control of the network. There are several innovative elements in tuning parameters of FASE-based, demand profiling, and aggregation. The comprehensive case studies for three UK representative demand scenarios in 2023, 2035, and 2050 demonstrated the effectiveness of the proposed approach. 
\end{abstract}

\begin{IEEEkeywords}
Active Distribution Networks, Energy Demand Forecasting, FASE, Synthetic Data Generation, and WaveNet-LSTM.
\end{IEEEkeywords}

\section{Introduction}
\IEEEPARstart{T}{he} integration of distributed energy resources (DERs) and advanced power electronically controlled loads in distribution networks (DNs) present several challenges, mainly due to their high variabilities and associated uncertainties~\cite{ref1}. Network operations like volt-var control, utility operational planning, and protection rely on accurate knowledge of the operational quantities, typically obtained through power system state estimation with the help of measurements through SCADA systems and more recently through Phasor Measurement Units (PMUs)~\cite{ref43}. Even though Advanced Metering Infrastructures (AMIs) and smart meters are penetrating in distribution networks, these measurement sources might not be available for distribution networks due to communication constraints and privacy concerns.

Increasing penetrations of DERs bring new challenges for distribution system state estimation mainly because the static estimation techniques which were primarily developed for passive networks can no longer be applied to active distribution networks. The dynamics introduced by inverter-based resources/DERs are spatially uneven and vary vastly in timescales, making it difficult to apply purely static or dynamic state estimation methods to monitor the entire network. In addition, the uncertainties in DER dynamics further complicate the state estimation task~\cite{ref2}. To address these challenges, combined dynamical and static state estimation methods need to be developed which include a distribution system model consisting of differential and algebraic equations to fully capture the dynamics at different timescales. This paper seeks to elaborate on the urgent need for new estimation methods and discuss potential solutions, including FASE based on online-tuning smoothing parameters and Extended Kalman Filter (EKF).

\par There are many approaches proposed in the literature for distribution network monitoring including static state estimation such as Weighted Least Squares (WLS), Weighted Least Absolute Value (WLAV), Schweppe-type Huber generalized maximum-likelihood (SHGM), and Forecasting-Aided State Estimation (FASE) algorithms based on Kalman Filter, etc. Static state estimators are widely used which minimize the errors between the estimated states and the pseudo measurements, but struggle to handle the uncertainties of the load demands and unmeasured and variable renewable energy generation~\cite{ref36}. Also,
FASE based on Kalman Filter can provide better accuracy compared to the static state estimation algorithms that depend on minimizing the error between the measurements and the estimated states since it uses a time-varying model to estimate the states~\cite{ref4}. Furthermore, load variations between static points can increase the error since these static state estimation methods rely on fixed snapshots of demands. 
\par The concept of FASE has been studied by trying to capture the dynamic behavior of the power systems states using Kalman Filter~\cite{ref5,ref6}. The use of Artificial Neural Network (ANN) with the FASE was proposed in~\cite{ref7}. They forecasted the bus load to enhance the accuracy of the estimated states of a distribution network. 

\par One of the challenges in DSSE is the lack of observability. A network system is called observable when the distribution system’s states can be accurately known based on the available measurements. The states of the DNs cannot be estimated unless the set of measurements cohesively describes each node. Most distribution networks rarely have a whole set of real-time measurements that describe each node; therefore, the system becomes under-determined. Additional information that can improve the observability are pseudo measurements and virtual measurements. Pseudo-measurements can be extracted by various techniques such as statistical approaches~\cite{ref46}, load forecasts, or using advanced metering infrastructure~\cite{ref8}. Smart meters have been used to generate pseudo measurements. In~\cite{ref9}, the pseudo load profile in low voltage DNs was determined using billing cycles of energy consumption and a few customers with smart meters. However, these measurements from smart meter units are not always synchronized. Thus, having visibility of the distribution network power demands is always a challenge.

\par The development of a suitable DSSE for ADNs requires a dataset that describes the behavior of the electricity consumption. Recent work introduced a new method to create detailed electricity demand profiles from consumption data using Empirical Mode Decomposition (EMD) compared against the Standard Load Profiles (SLPs), which were mostly measured in the 60s and 70s~\cite{ref10}. However, these SLPs-based methods may not capture all influencing factors on the electricity consumption such as the behavior of the current devices. On the other hand, the EMD method can be computationally intensive and depends on the availability of high-resolution data. Obtaining quality data is tough due to the cost and privacy concerns, emphasizing the need for synthetic data~\cite{ref11}. Challenges remain in extending these models to various customer types, making them adaptable, and ensuring replicability~\cite{ref12}. Continued research is vital to improve load profile modeling and data generation in DER-dominated power distribution systems.

\par In this paper, we propose a structured approach that addresses these challenges of lack of visibility and enhancing the performance of FASE based on Extended Kalman Filter (EKF). 

The proposed approach includes low-side demand profiles containing residential loads including DERs sources, time of use price for the home battery energy storage system (BESS), and small and medium enterprise's (SME) demand profiles. The realistic-synthetic distribution network power demands represent the complex structure and high level of irregularities of electricity consumption. 

The FASE based on Holt's Linear Method~\cite{ref3} smoothing parameters are adaptively tuned according to the main branch flow current and analysis through the proposed FASE-based DSSE. In~\cite{ref13}, the smoothing parameters are tuned using RMSE for a one-time sample which is time-consuming. We are obtaining online dynamic smoothing parameters for an unbalanced three-phase distribution network to enhance the performance of the state estimator. A hybrid WaveNet and Long Short-Term Memory(WaveNet-LSTM) forecaster is developed to forecast the Medium Voltage (MV) load profile based on the weather forecast, real-time measurements, and the aggregated data of some available smart meters. 


  
  

Section II introduces the problem formulation, Section III presents case studies, followed by a summary and conclusion in the sequel.


\section{PROBLEM FORMULATION}
\vspace{5pt}
The proposed WaveNet-LSTM model is developed because of its ability to capture the complex patterns of power demand, considering elements such as Photovoltaics (PV) systems, Heating, Ventilation, and Air Conditioning systems (HVAC). Also, a key part of this structured approach is refining the FASE algorithm by determining the adaptive values of tuning smoothing parameters. This enhances the accuracy and adaptability of our demand predictions within an ADN and state estimation process. For this, a synthetic DN demand profile representing the behavior of ADNs has been introduced.
\subsection{\bfseries WaveNet-LSTM Regrssion Model}
\vspace{5pt}
The proposed model combines the strengths of WaveNet and LSTM architectures, where it can capture complex patterns in time series data with high levels of irregularities. Specifically, the complex patterns and the long-range dependencies of the power demand profiles. While WaveNet can capture complex patterns in data because of the dilated convolutional layers, LSTM is effective in capturing temporal dependencies that are imposed by the dynamics of power demand profiles. 

\vspace{5pt}
\subsubsection{\bfseries Time Series Data}
\vspace{5pt}
\par Before training the developed WaveNet-LSTM model, the time series data needed to be transformed into a supervised learning format~\cite{ref31}. This was done by converting the data into a matrix of input and output pairs. The past observations are converted as a sequence that shapes the input whereas the output is shaped by the observation at the next time step. This is obtained by creating a sliding window of fixed size throughout the time series data. Each window shifted by one time step.

After creating the sliding window, the data was split into training, validation, and test sets that had not been seen by the model ‘holdout data’. The model fitted with the training data. The training of the model was completed when the model could predict future observations using the test dataset. This was done by choosing a suitable algorithm that suits the characteristics of the dataset and identifying the hyperparameter of the model. Various algorithms that can have a good ability to handle time series data with high levels of irregularities are Long Short-Term Memory (LSTM)~\cite{ref30}, and CNN-LSTM, however, the proposed method WaveNet-LSTM in demand forecasting outperformed them.
\vspace{5pt}
\subsubsection{\bfseries WaveNet-LSTM Architecture}
\vspace{5pt}
\begin{figure*}[t]
\centering
\includegraphics[width=\textwidth]{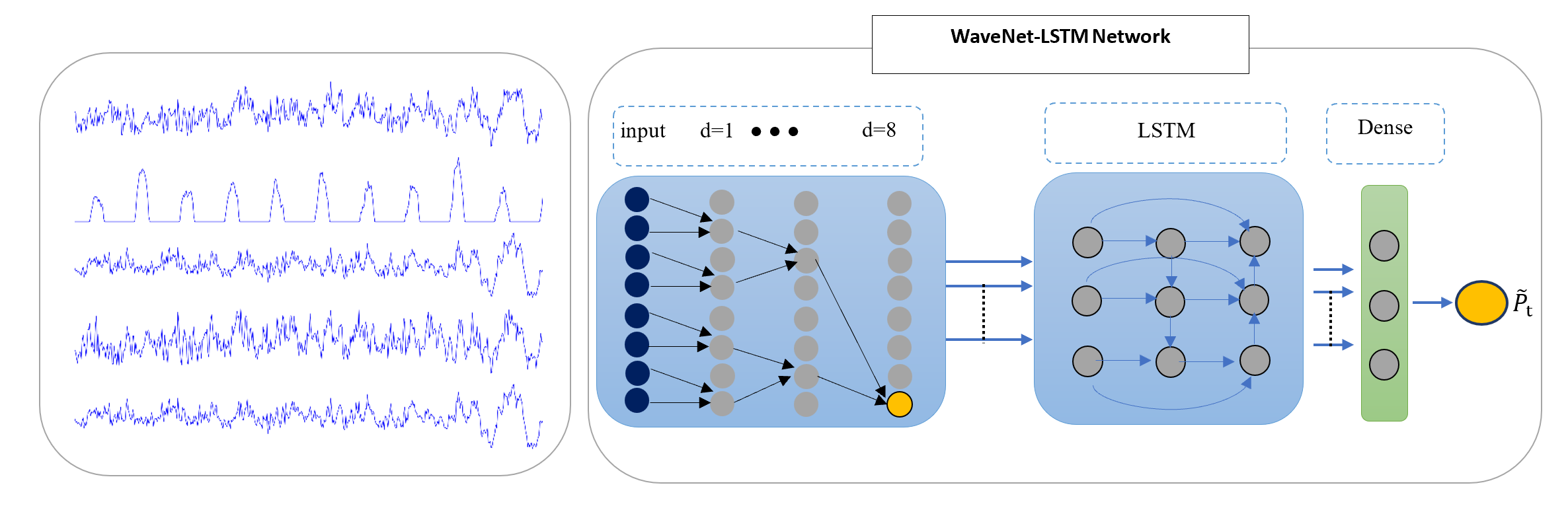}
\caption{Block diagram of proposed WaveNet-LSTM.}
\label{fig:load_forecaster}
\end{figure*}
\par The WaveNet-LSTM model is considered a multi-stage approach, as shown in Fig.~\ref{fig:load_forecaster}, to extract and learn sophisticated temporal patterns within complex data. The model was created based on WaveNet dilated convolutional layers, where skip connections play a vital role in capturing intricate features~\cite{ref35}. These skip connections create a deep hierarchy that effectively integrates the output of convolutional layers. This cascade of features is then further refined by the LSTM layers, which can capture long-term dependencies~\cite{ref34}. These LSTM units use a leaky ReLU activation function and dropout regularization to ensure stable learning~\cite{ref32,ref33}.The activation function is defined as:
\begin{equation}\label{eq:activation}
f(x)=
\begin{cases}
\rho x, & \text{if } x < 0 \\
x, & \text{if } x \geq 0
\end{cases}
\end{equation}
where $\rho=0.03$ represents a small gradient when the unit is not active.

The architecture of the WaveNet-LSTM model exhibits a similar capacity for adaptability and customization, allowing adjustments based on the properties and parameters of the constituent layers. The WaveNet-LSTM model incorporates WaveNet dilated convolutional layers, LSTM units, dropout regularization, and output layers. Each of these elements can be tuned to enhance performance in line with the characteristics of the data being modeled, in this case, MV demand power.

This model is customized to allow adjustments to the layers, such as parameters like the number of filters, kernel sizes, and dilation rates within the dilated convolutional layers to capture demand power trends over time. The WaveNet dilated convolutional layers are used to capture temporal features within these 30-minute intervals of power demand. Once the convolutional layers have processed the data, the LSTM units engage, leveraging their sequential nature to capture sequential dependencies. Finally, a dense layer smoothens out the output with a linear activation function. The hybrid architecture exhibits remarkable performance, ensuring that the model can efficiently interpret and forecast demand power dynamics at this interval. The hyperparameters of the proposed model are shown in Table~\ref{tab:hyperparameters}.

\begin{figure}[hb!]
    \centering
    \includegraphics[width=8.5cm,height=4.8cm,keepaspectratio]{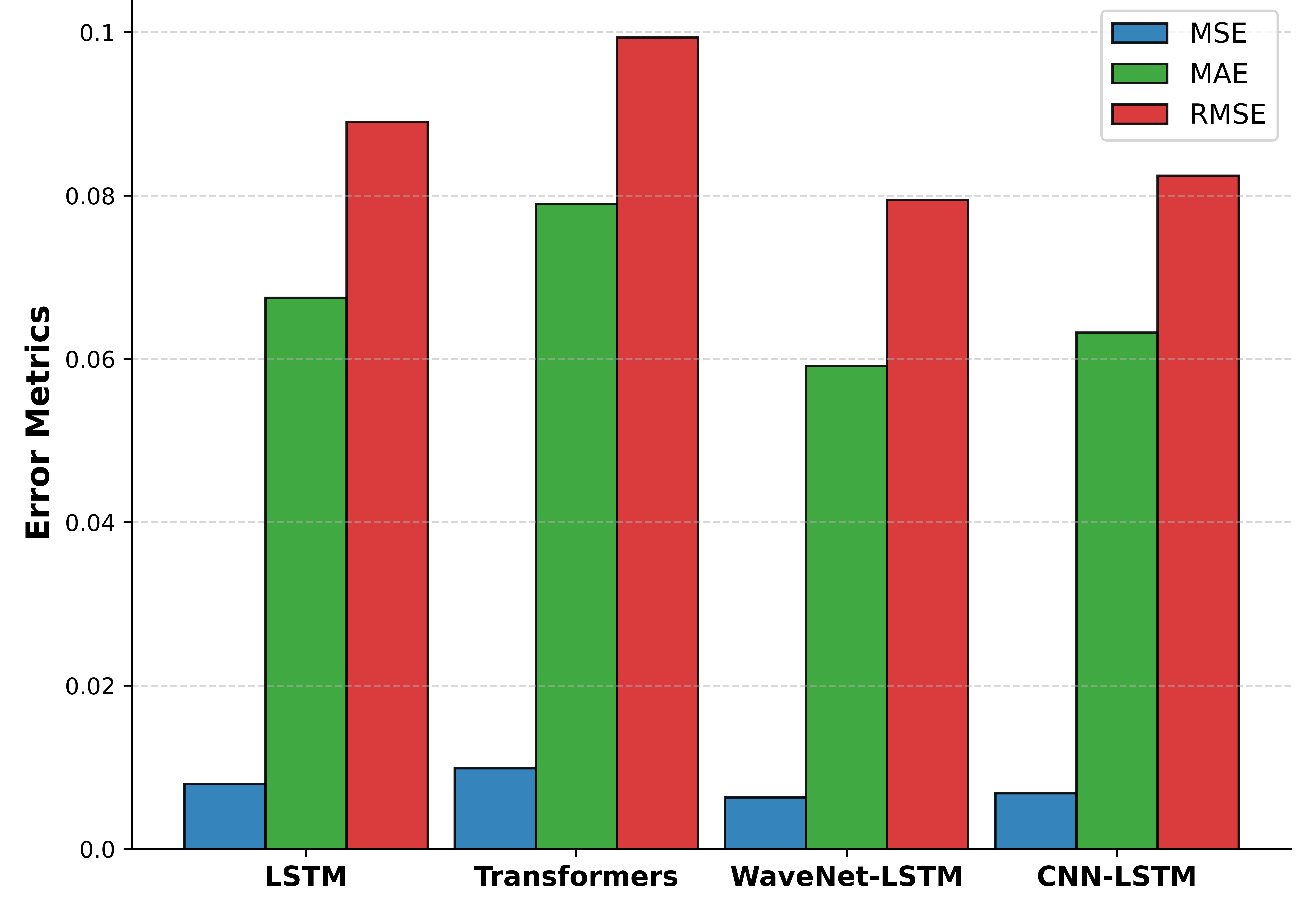}
    \caption{Performance comparison for different models.}
    \label{fig:metric}
\end{figure}
\vspace{-7pt}
\subsubsection{\bfseries WaveNet-LSTM Evaluation }
\vspace{5pt}
\par To validate the efficiency of the WaveNet-LSTM power demand prediction model introduced in this study, comprehensive experiments were conducted comparing various machine learning algorithms such as LSTM, Transformers Neural Network, and CNN-LSTM. Notably, the proposed WaveNet-LSTM model exhibited the lowest mean squared error (MSE) when compared against state-of-the-art machine learning techniques used for time series regression as shown in Table~\ref{tab:model_comparison} and Fig.~\ref{fig:metric}. The comparison was carried out utilizing the normalized values of MSE, MAE, and RMSE~\cite{ref34}.




\begin{table}
    \centering
    \caption{Performance Comparison of Models.}
    \label{tab:model_comparison}
    \begin{tabular}{lcccc}
        \toprule
        \textbf{Model} & \textbf{MSE} & \textbf{MAE} & \textbf{RMSE} \\
        \midrule
        LSTM & 0.0079 & 0.067 & 0.089 \\
        CNN-LSTM & 0.0067 & 0.063 & 0.082 \\
        TNN & 0.009 & 0.079 & 0.099 \\
        WaveNet-LSTM & \textbf{0.006} & \textbf{0.059} & \textbf{0.079} \\
        \bottomrule
    \end{tabular}
\end{table}

\begin{table}
\centering
\caption{Hyperparameters for the WaveNet-LSTM Model.}
\label{tab:hyperparameters}
\begin{tabular}{ll}
\toprule
\textbf{Hyperparameter}          & \textbf{Value}      \\
\midrule
Number of Dilation Layers        & 2                   \\
Number of Filters                & 32                  \\
Number of LSTM Layers            & 3                   \\
Number of Dense Layers           & 1                   \\
LSTM Units                       & 80                  \\
LSTM Dropout Rate                & 0.2                 \\
Learning Rate                    & 0.001               \\
Batch Size                       & 64                  \\
Loss Function                    & MAE                 \\
Optimizer                        & Adam LR=0.001       \\
Early Stopping Patience          & 10                  \\
\bottomrule
\end{tabular}
\end{table}
\vspace{-3pt}
\subsection{\bfseries Demand Profiles }
\vspace{5pt}
In the context of DSSE, an important aspect lies in generating synthetic demand profiles for ADNs to facilitate more realistic studies. These synthetic demand profiles utilize the projected energy scenarios outlined by the National Grid Electricity System Operator (ESO)~\cite{ref18} that have considerable impacts on future energy demands through heat pumps, Electric Vehicles (EVs), PV systems, and BESS. The generation of synthetic demand profiles has been carried out including DERs, EVs, HVAC systems, and variations in residential and commercial profiles (including PV and BESS integration). The generic demand profiles at Low Voltage (LV) were aggregated to represent the MV distribution network power demands. Generic DER profile modeling, residential and commercial profiles, and aggregated demands for MV DN power demands are presented next. 
\vspace{5pt}
\subsubsection{\textbf{Generic DERs Profiles Modelling}}
\vspace{5pt}
\par Developing the behavior of generic DERs can provide an understanding of ADN including EV, HVAC, PV, and BESS.  
\vspace{-5pt}
\subsubsubsection{\textbf{Electric Vehicle}}
\mbox \\
\vspace{3pt}
\par Electric Vehicle (EVs) modeling must consider user behaviors, state of charge (SOC) dynamics, and charging durations. Utilizing real-world driving survey data helps in capturing driving patterns and charging behavior as well as various variables such as travel distances, charging frequencies, and energy consumption profiles. User choices and vehicle specifications were addressed by considering various EV brands~\cite{ref14} with variable battery capacities~\cite{ref15} reflecting behaviors of 219,194 users like plug-in time, SOC upon return, and charging duration calculations. 
The initial state of charge (\(SOC_i\)) is computed as~\cite{ref16}:
\begin{align}
SOC_i = 1 - \frac{E_{dr}}{C_{\text{batt}}}
\end{align}

The charging duration (\(Charging_{\text{dur}}\)) can be calculated as\cite{ref17}:
\begin{equation}
Charging_{\text{dur}} = \frac{C_{\text{batt}} \left(1 - SOC_i\right)}{P_{\text{rated}} \cdot \eta} \quad \forall  SOC_i > 0
\end{equation}

Where \(C_{\text{batt}}\) is the battery capacity in kilowatt-hours (kWh), \(P_{\text{rated}}\) is the power rating of the charger in kilowatts (kW), which is determined by the charger level (residential level in this case), \(E_{\text{dr}}\) represents the energy discharged from the battery, and \(\eta\) is the efficiency of the charger expressed as a percentage.

Additionally, we integrate a charging dynamic model outlined in~\cite{ref2}, characterized by the following equations:

\begin{equation}
RC(t+1) = RC(t) - u(t)
\end{equation}
\begin{equation}
E(t) = P_{\text{rated}} \cdot u(t)
\end{equation}

Where \(RC\) is the remaining time to complete the charging task, and \(u\) is the control variable that specifies the action of charging, where \(u=1\) is for charging and \(0\) is for not charging.

We created individualized Electric Vehicle (EV) demand profiles utilizing insights from the behaviors of 219,194 users through the National Household Travel Survey (NHTS) and the consideration of the mentioned 12 EV brands. These individualized EV demand profiles were then aggregated and averaged.
\vspace{-8pt}
\subsubsubsection{\textbf{HVAC System}}
\mbox \\
\vspace{5pt}
\par The Heating, Ventilation, and Air Conditioning (HVAC) systems, including heat pumps, play a vital role in shaping the future energy scenarios in the UK~\cite{ref18}. The HVAC system was modeled as an RCQ circuit~\cite{ref19}. The dynamic model of the HVAC systems is described using a state-space representation:

\begin{equation}
\dot{x} = Ax + Bu
\end{equation}
\begin{equation}
y = Cx + Du
\end{equation}

Here, the matrices are defined as follows:
\[
A = \begin{bmatrix} -\left(\frac{1}{R_2C_a} + \frac{1}{R_1C_a}\right) & \frac{1}{R_2C_a} \\ \frac{1}{R_2C_a} & -\frac{1}{R_2C_m} \end{bmatrix}
\]
\[
B = \begin{bmatrix} \frac{T_0}{R_1C_a} + \frac{Q}{C_a} \\ 0 \end{bmatrix}
\]
\[
C = \begin{bmatrix} 10 & 0 & 1 \end{bmatrix}, \quad D = \begin{bmatrix} 0 \end{bmatrix}
\]
In the above expressions:
\begin{align*}
& C_a: \text{Air heat capacity (}J/^\circ C\text{)}. \\
& C_m: \text{Mass heat capacity (}J/^\circ C\text{)}. \\
& Q: \text{Heat rate for HVAC unit (}W\text{)}. \\
& UA: \text{Standby heat loss coefficient of wall (}W/^\circ C\text{)}. \\
& UA_{\text{mass}}: \text{Heat loss coefficient of thermal mass (}W/^\circ C\text{)}. \\
& R_1: \frac{1}{U_A}. \\
& R_2: \frac{1}{UA_{\text{mass}}}. \\
& T_0: \text{Ambient temperature (}^\circ C\text{)}. \\
& T_i: \text{Air temperature inside the house (}^\circ C\text{)}. \\
& T_m: \text{Mass temperature inside the house (}^\circ C\text{)}.
\end{align*}

In the thermal equivalent circuit or RCQ model, the resistances represent heat loss through walls and windows, while the capacitor represents the thermal mass created by the internal room temperature and furniture. During heating mode, electricity injection generates a positive heat rate (Q), while cooling involves a negative Q due to heat extraction by the compressor. The dynamics of the HVAC unit are effectively captured by a state-space model~\cite{ref19}. It is to be noted that the one-year generic HVAC system demand profile was generated using the average ambient temperature of the UK's weather over 14 years~\cite{ref21}, and the HVAC system parameters were obtained from~\cite{ref44,ref45}. Subsequently, the output was normalized based on the heat rate (Q) of the HVAC.

\vspace{-8pt}
\subsubsubsection{\textbf{Photovoltaic System}}
\mbox \\
\vspace{5pt}
\par In this study, we adopted a single-diode circuit of a PV cell which is modeled as a function of solar irradiance and the ambient temperature. The parameters of the PV module were obtained from Kyocera 200W~\cite{ref20}. The profiles were normalized to a maximum output of 200W. It is to be noted that the one-year weather data were obtained by averaging the UK weather for 14 years~\cite{ref21}.

The resultant output current, \(I_{PV}\), is intricately linked to the intensity of incoming light and is defined by the equation~\cite{ref22}:
\begin{equation}
I_{PV} = I_{ph} - I_D - \frac{V + IR_s}{R_sh}
\end{equation}

The calculation of diode current, \(I_D\), adheres to the equation:
\begin{equation}
I_D = I_o \left(e^{\frac{q(V + IR_s)}{nkT} - 1}\right)
\end{equation}

Here, \(I_o\) represents the diode's saturation current, \(T\) symbolizes the temperature of the p-n junction in Kelvin, while \(q\), \(n\), and \(k\) denote fundamental constants. The photocurrent, \(I_{ph}\), is derived from:
\begin{equation}
I_{ph} = I_{ph}(T_1) + K_o(T - T_1)
\end{equation}

Where \(T_1\) is the reference temperature. Constants \(K_o\) and \(I_L\) are determined as:
\begin{equation}
K_o = \frac{I_{SC}T_{1} - I_{SC}T_1}{T - T_{1}}
\end{equation}
\begin{equation}
I_L(T_1) = I_{SC}T_1 \left(\frac{G}{G_{nom}}\right)
\end{equation}

\(G\) represents solar irradiance, while \(G_{nom}\) signifies solar irradiance under standard conditions. Employing MATLAB, the PV module was intricately modeled using data from~\cite{ref20}. 
\vspace{-15pt}
\subsubsubsection{\textbf {Battery Energy Storage System}}
\mbox \\
\vspace{5pt}
\par Battery Energy Storage Systems (BESS) at the residential level play a vital role in shaping the future energy scenarios in the UK. The adoption of these technologies is predicted to be on the rise~\cite{ref18}. The BESS model in this study is based on the assumption that homes would have smart energy management systems, taking into account the time-pricing signal as well as the output from the rooftop PV system through deterministic linear programming~\cite{ref23,ref24}.

The core of this optimization revolves around the interplay of parameters, notably the relationship between PV power (\(S_t\)), grid power (\(g_t\)), battery power (\(b_t\)), load power (\(l\)), the battery capacity (\(e_t\)), and energy price (\(c_t\)). The mathematical expressions can be described as:

\begin{equation}
S_t + g_t = b_t + l
\end{equation}
\begin{equation}
e_{t+1} = e_t + b_t
\end{equation}

The objective of this optimization is to minimize the electricity bill, excluding the State of Health (SOH) of BESS for the sake of simplicity, subject to the following constraints:

\begin{equation}
\min \sum_{t} c_t \cdot g_t
\label{eq:your_label_here}
\end{equation}
\[
0 \leq e_t \leq e_{\text{max}}
\]
\[
|b_t| \leq b_{\text{max}}
\]
\[
e_0 = e_{-1}
\]

\subsubsection{\bfseries Residential and Commercial Demand Profile}
\vspace{5pt}
\par
\begin{figure}[t]
  \centering
  \includegraphics[width=8cm,height=5.5cm,keepaspectratio]{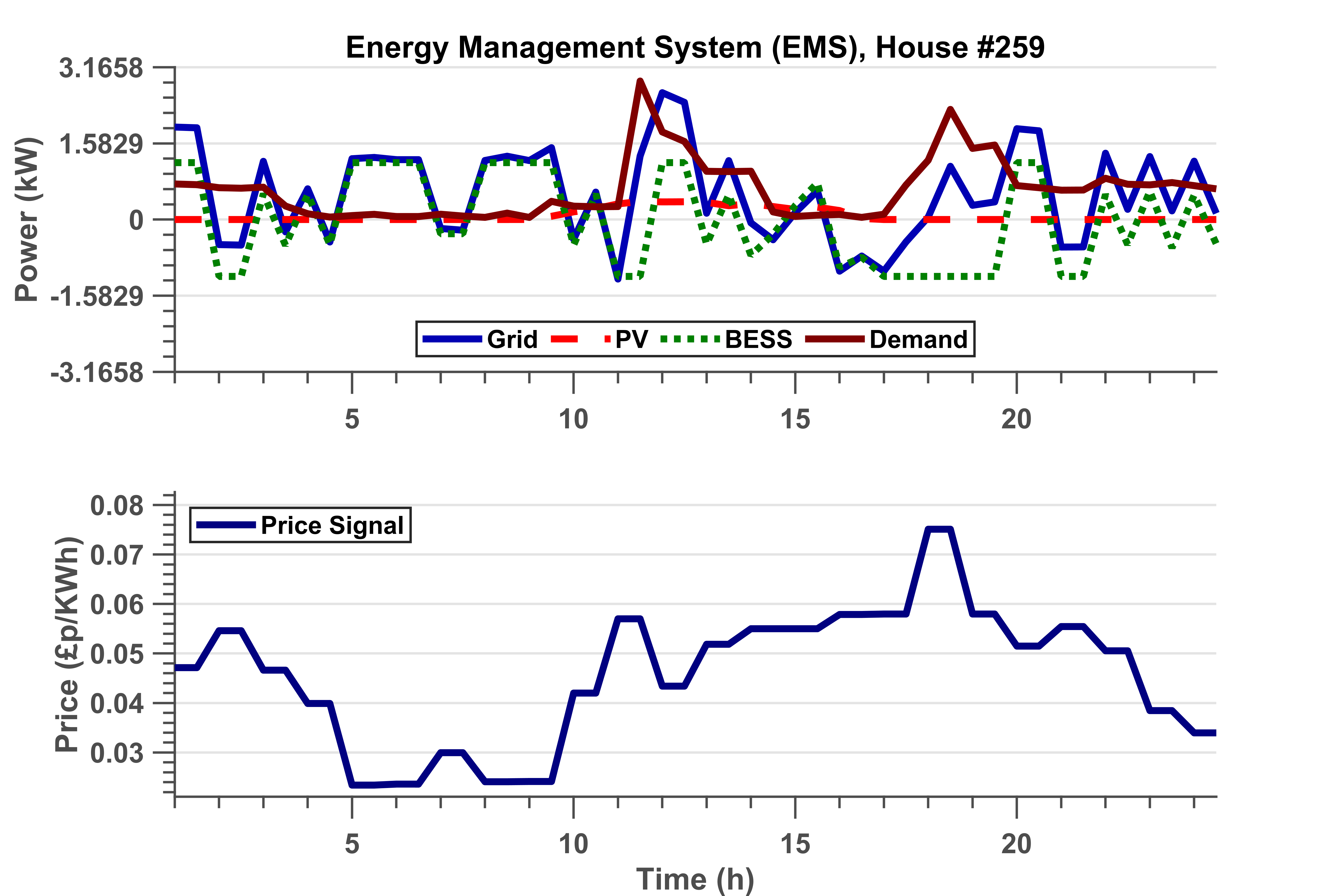} 
  \caption{Energy management system in action}
  \label{fig:ems}
\end{figure}

 Diverse demand profiles within a low-voltage distribution network pose a challenge within the context of DSSE. In this paper, 900 smart meter profiles that represent the energy consumption of 900 UK houses were obtained from~\cite{ref25}. Then we created a one-year dataset that is structured into three core categories, each encompassing 300 demand profiles. The first category captures the fundamental essence of households without the impact of PV systems, EV charging, or BESS. The second represents 300 homes that have PV systems to reduce the reliance on the grid electricity and send the excess of solar-generated energy to the grid. Finally, households with solar rooftop systems and BESS with a smart energy management system to reduce the cost according to the price signal~\cite{ref29}. For instance, Fig.~\ref{fig:ems} shows the energy management system in action for house \#259. It is to be noted that the sizing of the PV system and BESS was based on the annual average household energy consumption. Furthermore, the energy exports to the grid are set to flexible connection agreements~\cite{ref26}. The commercial demand profiles were obtained from~\cite{ref25}. 

\vspace{8pt}
\subsubsection{\bfseries Aggregation of MV Demand Profiles}
\begin{figure}[h]
  \begin{subfigure}{0.45\textwidth}
    \centering
     \includegraphics[width=9cm,height=4.8cm,keepaspectratio]{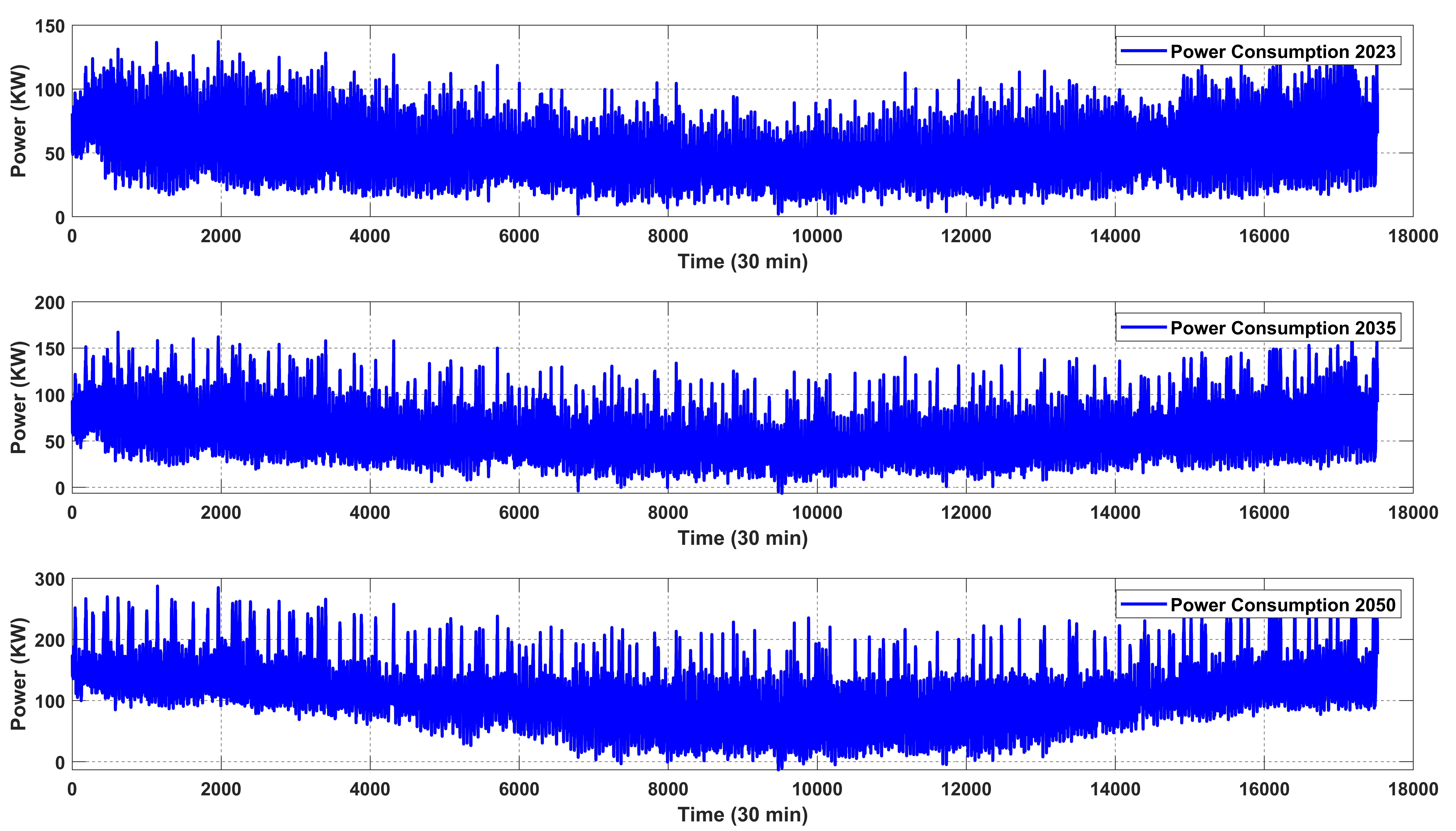}
    \caption{Annual demand consumption for three scenarios} 
    \label{fig:subfig1} 
  \end{subfigure}
  \begin{subfigure}{0.45\textwidth}
    \centering \includegraphics[width=8.5cm,height=4.8cm,keepaspectratio]{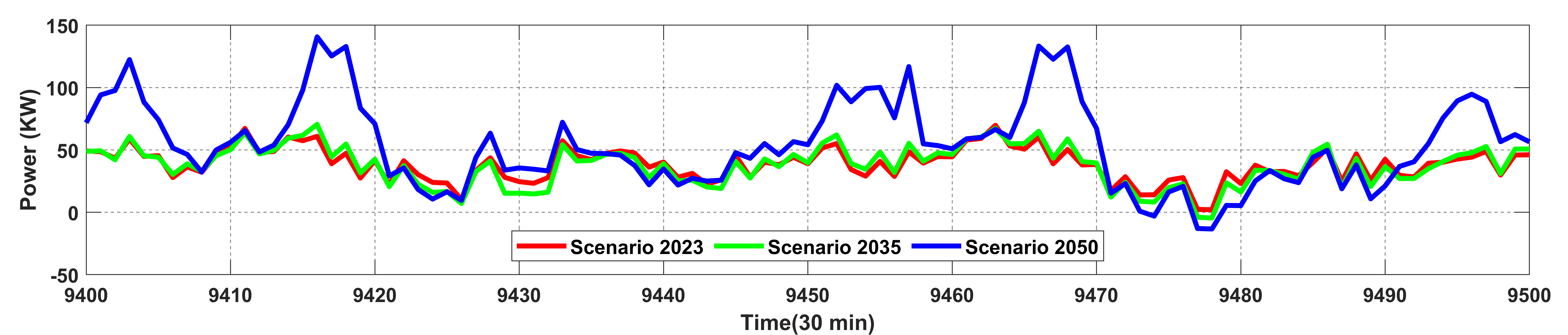} 
    \caption{Zoomed consumption for three scenarios from 9400 to 9500} 
    \label{fig:subfig2} 
  \end{subfigure}
  \caption{Synthetic UK power demands for different scenarios} 
  \label{fig:power-scenarios} 
\end{figure}

\begin{figure}[t]
    \centering
    \includegraphics[width=0.8\columnwidth]{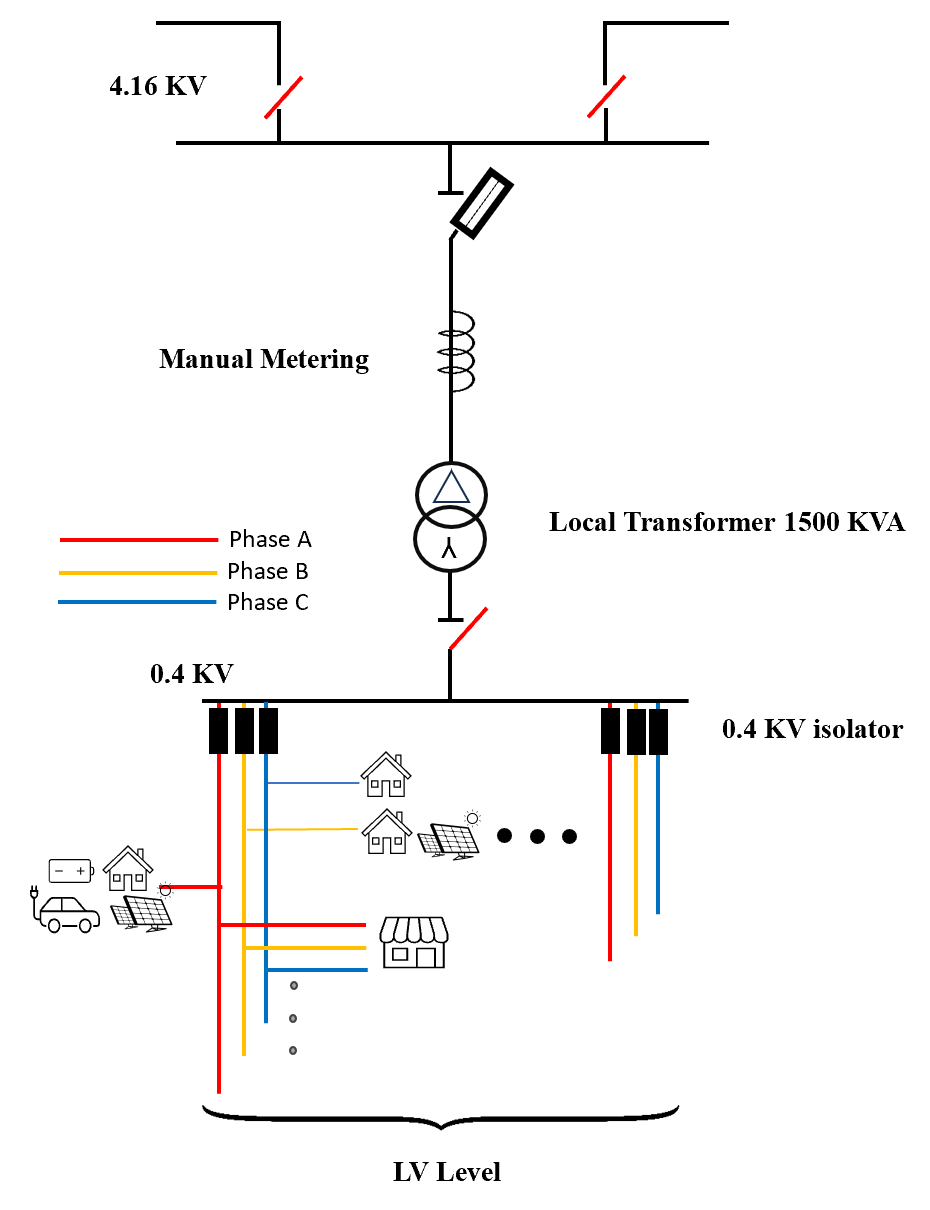}
    \caption{Layout of a feeder in a DN.}
    \label{fig:layout_DN}
\end{figure}

\vspace{5pt}
Constructing MV demand profiles involving the consideration of errors in the grid as well as realistic LV demand profiles is significant for having an accurate DSSE. The process of constructing MV demand profiles utilizes the created dataset of the downstream consumers, as shown in Fig.~\ref{fig:layout_DN}, considering three load compositions described in Table~\ref{tab:demand-profiles}. These compositions mimic the current UK energy scenario, and two predicted future energy scenarios for 2035 and 2050 using the statistics and probability of the Future Energy Scenarios report by ESO~\cite{ref18}. The process also takes into account different sources of errors in the MV demand aggregation. A random error was assumed to reflect factors such as phase identification, missing data errors, power losses, smart meter synchronization, and the accuracy of smart meter units~\cite{ref27}. Fig. ~\ref{fig:power-scenarios} illustrates an example of how the aggregated MV power demand can be different.


\begin{table*}[b]
  \centering
  \caption{Household and Commercial Demand Profiles.}
  \label{tab:demand-profiles}
  \resizebox{\textwidth}{!}{%
    \begin{tabular}{ccccccc}
      \toprule 
      \textbf{Year} & \textbf{Household Demand\%} & \textbf{Household Demand with PV\%} & \textbf{Household Demand with PV \& BESS\%} & \textbf{Commercial Profile \%} & \textbf{No. EV} & \textbf{No. HVAC} \\
      \midrule 
      2023 & 79 & 5 & 1 & 15 & 4 & 5 \\
      2035 & 65 & 18 & 2 & 15 & 38 & 8 \\
      2050 & 44 & 37 & 4 & 15 & 80 & 40 \\
      \bottomrule 
    \end{tabular}}
\end{table*}
The aggregation process begins by merging the mentioned demand datasets, each corresponding to different load compositions and household profiles. These datasets encompass residential, commercial, HVAC, EVs, and other relevant loads, providing a holistic view of MV demand. At the core of the aggregation methodology lies the equation~\cite{ref28}:
\begin{equation}\label{eq:agg}
\begin{aligned}
\mathbf{P}_{\text{agg}} = \biggl[ & \biggl( \sum_{i=1}^{h} P_i(t) \biggr)_{t=1}, \biggl( \sum_{i=1}^{h} P_i(t) \biggr)_{t=2}, \\
& \ldots, \biggl( \sum_{i=1}^{h} P_i(t) \biggr)_{t=48} \biggr]
\end{aligned}
\end{equation}
Equation (\ref{eq:agg}) captures the essence of combining individual household-level power profiles \(p_i\) for each time slot \(t\) throughout the day, aggregated over \(h\) households. This summation generates the aggregated active power profile \(\mathbf{P}_{\text{agg}}\) for the entire MV system, relating consumption patterns over the 48 time intervals.

\subsection{\bfseries Forecasting Aided State Estimation }
\vspace{5pt}
Forecasting-Aided State Estimation depends on Kalman Filtering. The state space representation of the system is given by~\cite{ref36}:
\begin{equation}
    x_{k+1} = F_k x_k + g_k + w_k \label{state_update}
\end{equation}
\begin{equation}
    z_k = h_k(x_k) + v_k \label{eq:observation}
\end{equation}

FASE consists of two stages: the prediction and the correction. From \eqref{state_update}, the transition matrix \(F_k\) is responsible for the determination of the state vectors for the next time instant, while the state trajectory vector \(g_k\) is responsible for defining the level and the trend for the system. The \(w_k\) is white Gaussian noise and the \(v_k\) is the Gaussian error vector. Optimizing \(F_k\) and \(g_k\) represents a significant step in the FASE process. These parameters are determined using the Holt Linear Exponential method~\cite{ref36}. Smoothing the dataset to define the trend is accomplished with two parameters, denoted as $\alpha$ and $\beta$.
\begin{equation}
    \tilde{x}_{k+1}^i = a_k^i + b_k^i 
\end{equation}
\begin{equation}
    a_k^i = \alpha_i x_k^i + (1 - \alpha_i) \tilde{x}_k^i 
\end{equation}
\begin{equation}
    b_k^i = \beta_i (a_k^i - a_{k-1}^i) + (1 - \beta_i) b_k^i 
\end{equation}

Rewritten as:
\begin{equation}
    \tilde{x}_{k+1}^i = F_k^i \tilde{x}_k^i + g_k^i 
\end{equation}
\begin{equation}
    g_k^i = [(1+\beta_i)(1-\alpha_i)] \tilde{x}_k^i + \beta_i a_{k-1}^i + (1-\beta_i) b_{k-1}^i 
\end{equation}
Where:
\begin{equation}
    F_k^i = \alpha_i (1 + \beta_i)I 
\end{equation}
Where,
$\tilde{x}_k^i$: is the forecasted $i^\text{th}$ state vector. After conducting the forecasting stage, the state correction/filtering stage will take place to get the posterior state vector $\hat{x}_{k+1}^i$ and then the new error covariance $P_{k+1}$. This is done by minimizing the objective function in~\eqref{eq:objective} with the predicted state vector $\tilde{x}_{k+1}^i$ and the initial error covariance $P_{k+1}$~\cite{ref37}:

\begin{equation}
\begin{split}
    J(x) = &[z - h(\tilde{x}_{k+1}^i)]^T R^{-1} [z - h(\tilde{x}_{k+1}^i)] \\
    &+ (\hat{x}_{k+1}^i - \tilde{x}_{k+1}^i) P_{k+1}^{-1} (\hat{x}_{k+1}^i - \tilde{x}_{k+1}^i)^T
\end{split}
 \label{eq:objective}
\end{equation}

After the minimization, the Kalman gain is calculated as:

\begin{equation}
    K_{k+1} = P_k H_{k+1}^T R^{-1}
\end{equation}

So, the posterior estimate of the state vector and error covariance matrix is:

\begin{equation}
    \hat{x}_{k+1} = \tilde{x}_{k+1} + K_{k+1} (z_{k+1} - H_{k+1} \tilde{x}_{k+1})
\end{equation}

Similar logic to the ADMM penalty parameter, the developed strategy of having a dynamic smoothing parameter to improve the performance of the estimator is adopted from~\cite{ref38}. It is mathematically expressed as:

\begin{equation}
    \beta_{k+1}^{\phi} = \begin{cases}
        \beta_k^{\phi} + \tau & \text{if } \Upsilon_k \leq \Upsilon_{k+1} \\
        \beta_k^{\phi} - \tau & \text{if } \Upsilon_k \geq \Upsilon_{k+1} \\
        \beta_k^{\phi} & \text{otherwise}
    \end{cases}
    \label{eq:beta_update_phi}
\end{equation}

\begin{equation}
    \alpha_{k+1}^{\phi} = \begin{cases}
        \alpha_k^{\phi} + \epsilon & \text{if } \Upsilon_k \leq \Upsilon_{k+1} \\
        \alpha_k^{\phi} - \epsilon & \text{if } \Upsilon_k \geq \Upsilon_{k+1} \\
        \alpha_k^{\phi} & \text{otherwise}
    \end{cases}
    \label{eq:alpha_update_phi}
\end{equation}

Subject to:
\begin{align}
    \alpha^{\phi} &> \beta^{\phi}
    \label{eq:constraint_phi}
\end{align}

Here, $\tau$ represents the value used to increment or decrement $\beta$, $\epsilon$ signifies the value used for modifying $\alpha$, and $\Upsilon$ is the threshold for rate changes in the main branch flow currents. Additionally, $\phi$ is utilized to indicate the phase. The values of  $\tau$ ,$\epsilon$, and $\Upsilon$ are shown in Table~\ref{tab:variable_values}.Since there is no closed form expression, we obtained these values through extensive simulations.

\begin{table}[h]
    \centering
    \caption{Values of $\tau$,$\epsilon$ $\&$ $\Upsilon$}
    \label{tab:variable_values}
    \begin{tabular}{lc}
        \toprule
        \textbf{Variable} & \textbf{Value} \\
        \midrule
        $\tau$ & 0.013 \\
        $\epsilon$ & 0.01 \\
        $\Upsilon$ & 0.09 \\
        \bottomrule
    \end{tabular}
\end{table}


\begin{figure*}[b]
  \begin{minipage}{0.5\textwidth}
    \centering
    \begin{subfigure}{\linewidth}
      \includegraphics[width=\linewidth]{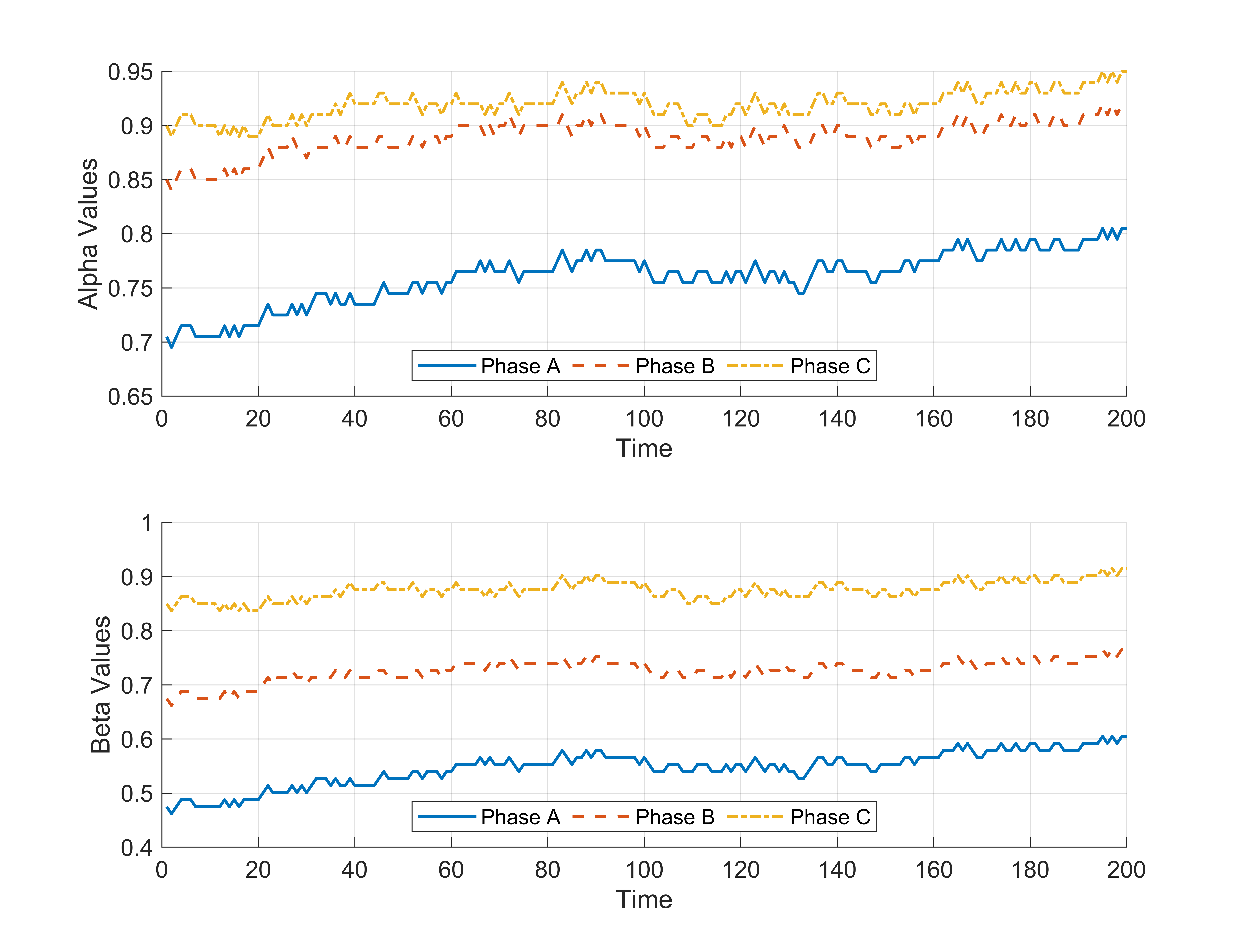}
      \caption{Alpha Beta adaptive values for FASE at scenario 2035 }
      \label{fig:alpha_beta_comparison}
    \end{subfigure}
  \end{minipage}%
  \begin{minipage}{0.5\textwidth}
    \centering
    \begin{subfigure}{\linewidth}
      \includegraphics[width=\linewidth]{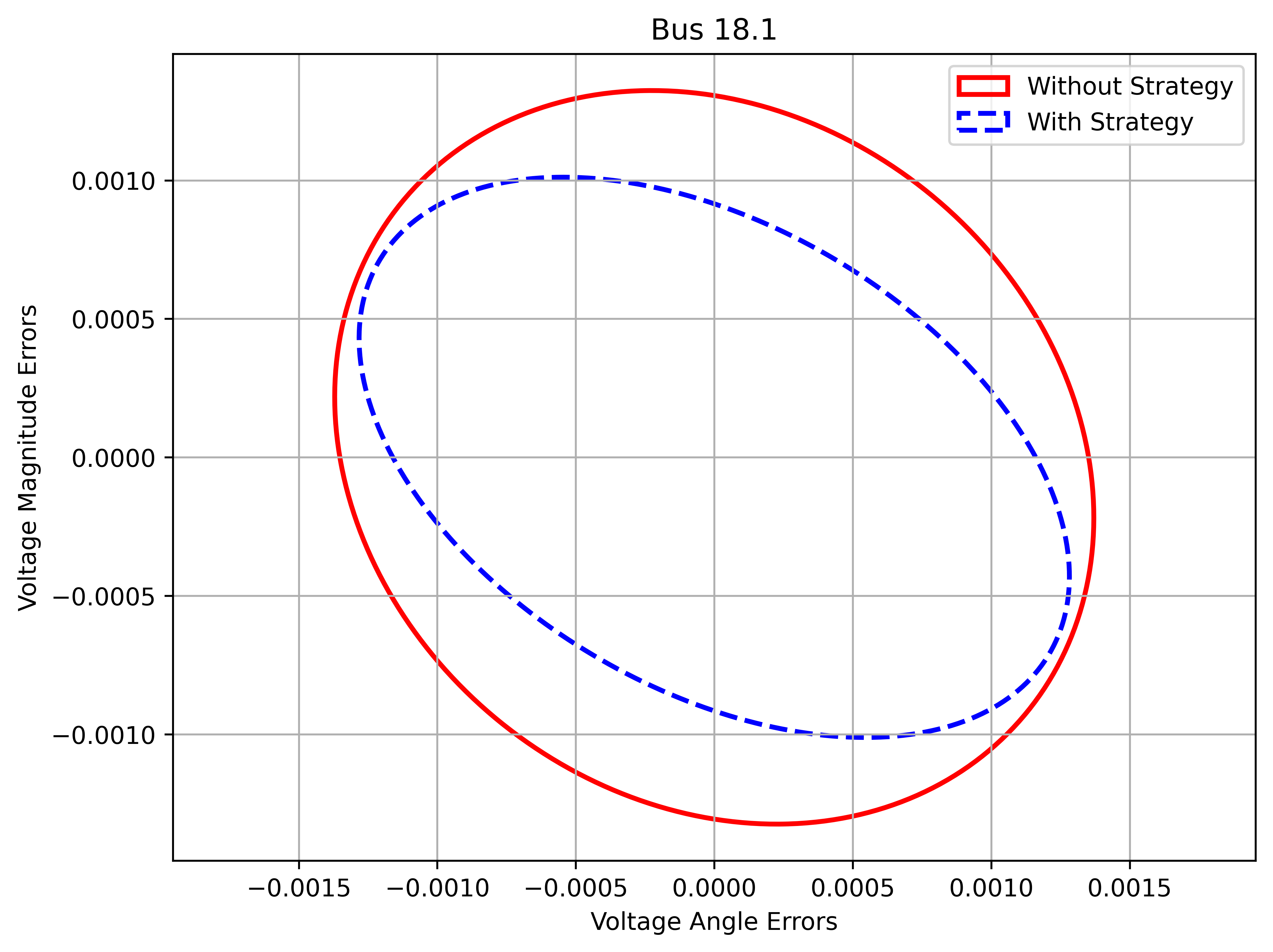}
      \caption{Error ellipses at bus18.1}
      \label{fig:bus18_updated3}
    \end{subfigure}
  \end{minipage}
    \caption{Dynamic $\alpha$ \& $\beta$ values results.}
    \label{fig:combined_plots}
  
\end{figure*}
\vspace{-5pt}
\section{Case Studies}
\vspace{5pt}
In this section, we comprehensively assess the viability of our proposed structured approach utilizing the realistic and synthetic data created in this study by three case scenarios rooted in the UK's power demand dynamics, including current demand, as well as prediction-based demands for 2035 and 2050. These case studies are conducted using a three-phase unbalanced IEEE 123-bus system~\cite{ref41}, a widely recognized benchmark comprising 197 power lines and various loads.


\vspace{-5pt}
\subsection{\bfseries Setup}
\vspace{5pt}
It was assumed that the unbalanced distribution network IEEE 123 test system has 18 real-time measurements including the flow powers and the three-phase voltage magnitude and phase angles at main transformer~\cite{ref40}\& \cite{ref2} as well as using the one main current sensor at the branch 8-13 in the test system. Also, to handle the issue of the measurement synchronization among smart meter units at the LV side, we assumed that at each time sample only approx. 40~\% of the smart meter units were available. The power consumption at the medium voltage side was predicted using the novel WaveNet-LSTM model created in this structured approach. To improve the capability of the MV power demands, we used solar irradiance as an extra feature in our power demand forecaster as well as the aggregation of some of the smart meter units, and the main branch flow currents as input features to have a better prediction of the MV power demands. It is to be noted that the input features must be compatible with the different three unbalanced phases. For instance, we need to use phase A measurements as input features when we are looking for the power demand value of bus 1 phase A. For the mentioned load compositions, we used the same locations that were used for DERL in~\cite{ref2}.

\begin{figure}[t]
    \centering
    \includegraphics[width=0.45\textwidth]{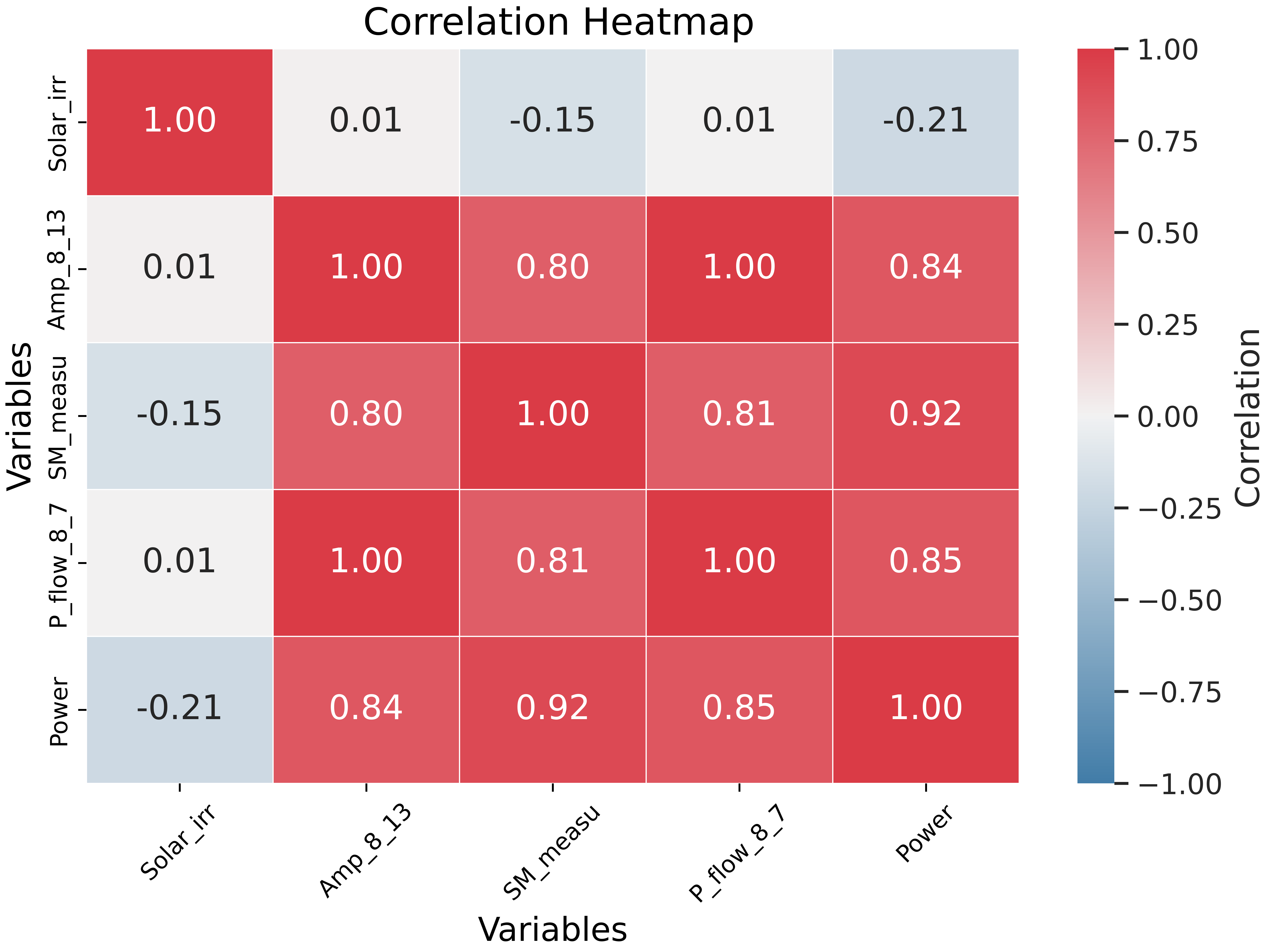}
    \caption{Correlation of the available features.}
    \label{fig:heatmap}
\end{figure}

 One of the challenges was that the IEEE benchmarks test systems come with fixed system loading conditions. As a solution, we generated synthetic and realistic power profiles as part of our research efforts to have a time-varying consumption profile for each node within the system. While these profiles may vary across nodes, we ensured that the annual average power consumption matched the snapshot power values provided by the IEEE 123 test system~\cite{ref2} then we normalized the obtained yearly time-varying consumption profile by dividing it by its maximum value. For the sake of simplicity, we focused on forecasting the real power profiles using the WaveNet-LSTM model while keeping the power factors at each node constant as provided in the IEEE 123-test system, where the time-varying reactive power profiles were derived from the forecasted real power profiles and the constant power factor. The correlation heatmap between the input features of the forecaster is shown in Fig.~\ref{fig:heatmap}.

Another challenge in this study was identifying the dynamic values of the smoothing parameters ($\alpha$ and $\beta$). This was addressed by following the strategy provided in the euqations~\ref{eq:beta_update_phi} $\&$ \ref{eq:alpha_update_phi}. Using the mentioned strategy has improved the estimated state, as shown in Fig.~\ref{fig:bus18_updated3}. 

The values of the smoothing parameters for the year 2035 are depicted in Fig.~\ref{fig:alpha_beta_comparison}. Finally, the performance of the proposed approach was assessed using the true values generated by BFS power flow analysis~\cite{ref39}.

\subsection{\bfseries Scenario 2023}
\begin{figure}[b]
\centering
  \includegraphics[width=\linewidth]{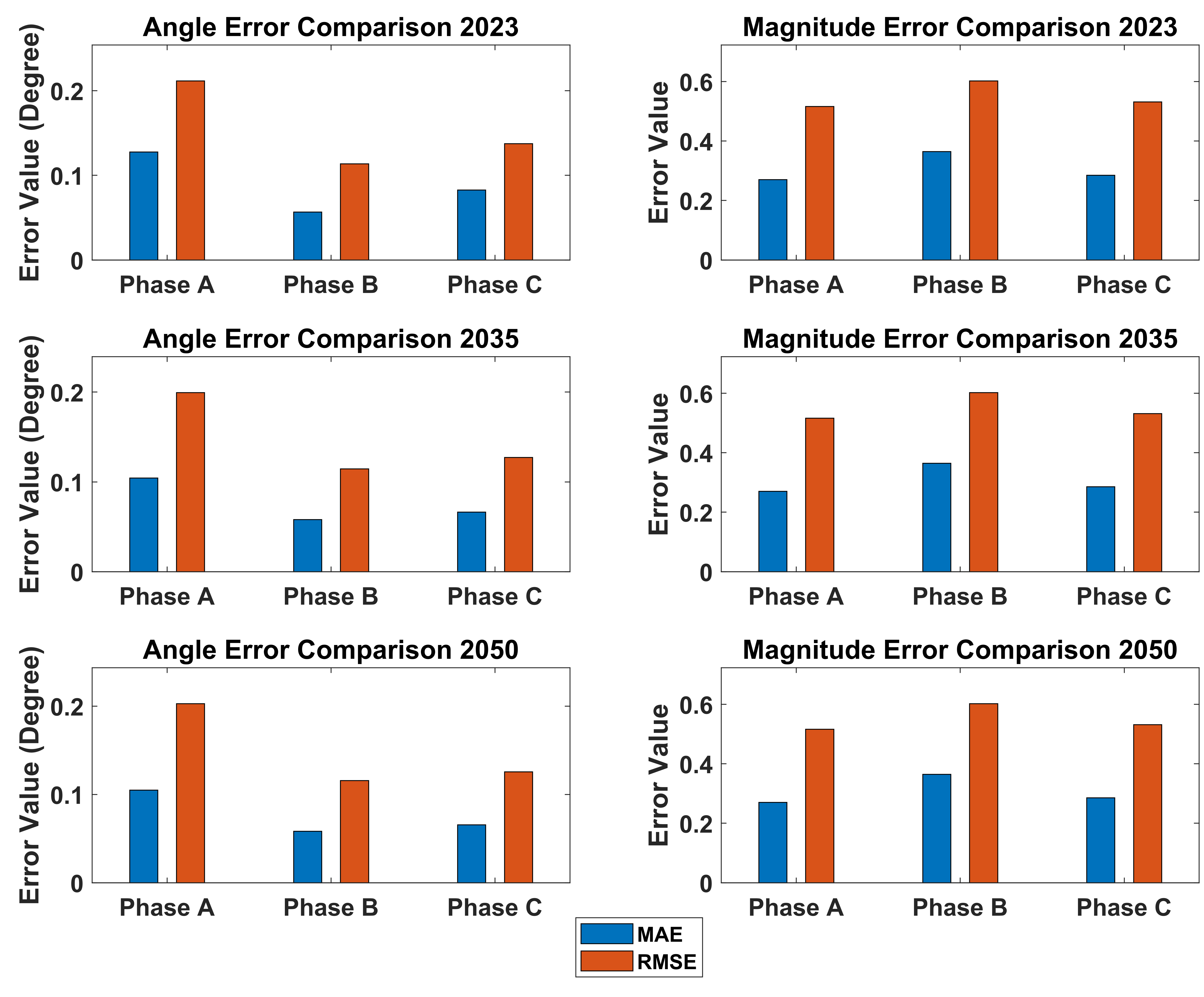}
    \caption{Error metrics for estimated states in 2023,2035,2050.}
    \label{fig:2023}
\end{figure}

\begin{figure*}[t]
  \begin{minipage}{0.48\textwidth}
    \centering
    \begin{subfigure}{\linewidth}
      \includegraphics[width=0.9\linewidth]{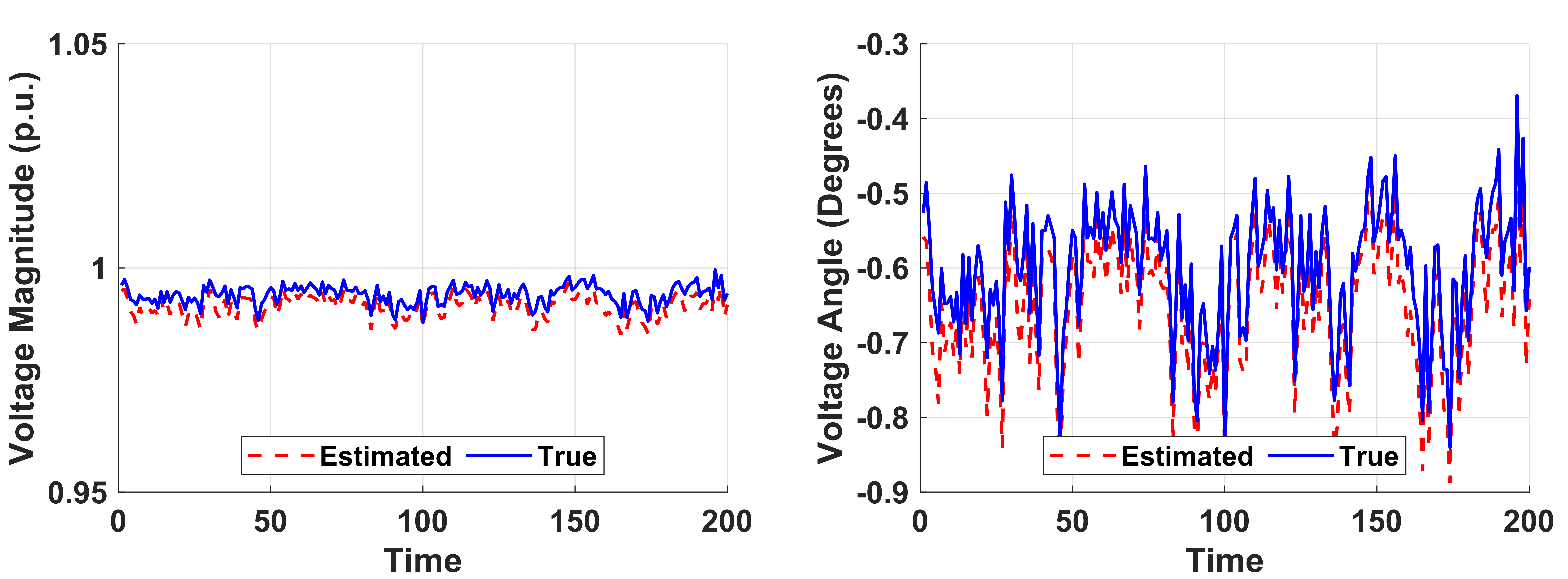}
      \caption{Voltage magnitude and angle at Bus 11.1 }
      \label{fig:subplot1}
    \end{subfigure}
  \end{minipage}%
  \begin{minipage}{0.48\textwidth}
    \centering
    \begin{subfigure}{\linewidth}
      \includegraphics[width=0.9\linewidth]{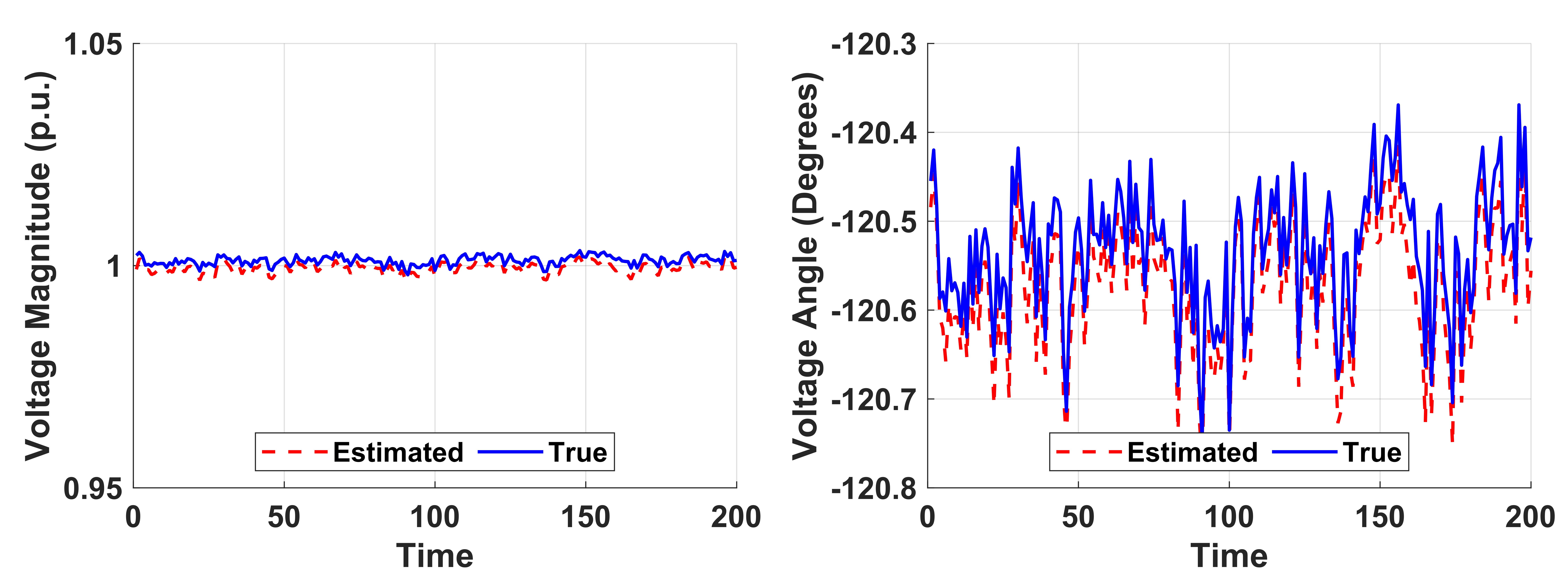}
      \caption{Voltage magnitude and angle at Bus 54.2 }
      \label{fig:subplot2}
    \end{subfigure}
  \end{minipage}
  
  \begin{minipage}{0.48\textwidth}
    \centering
    \begin{subfigure}{\linewidth}
      \includegraphics[width=0.9\linewidth]{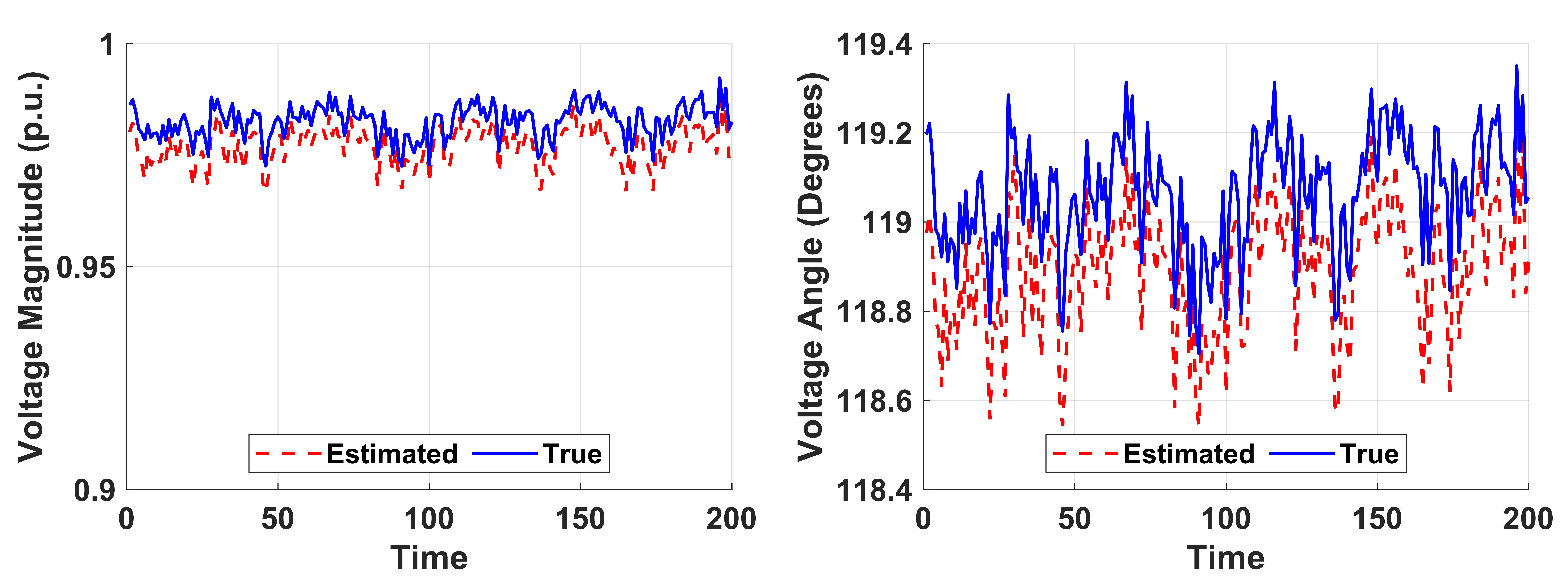}
      \caption{Voltage magnitude and angle at Bus 83.3}
      \label{fig:subplot3}
    \end{subfigure}
  \end{minipage}%
  \begin{minipage}{0.48\textwidth}
    \centering
    \begin{subfigure}{\linewidth}
      \includegraphics[width=0.9\linewidth]{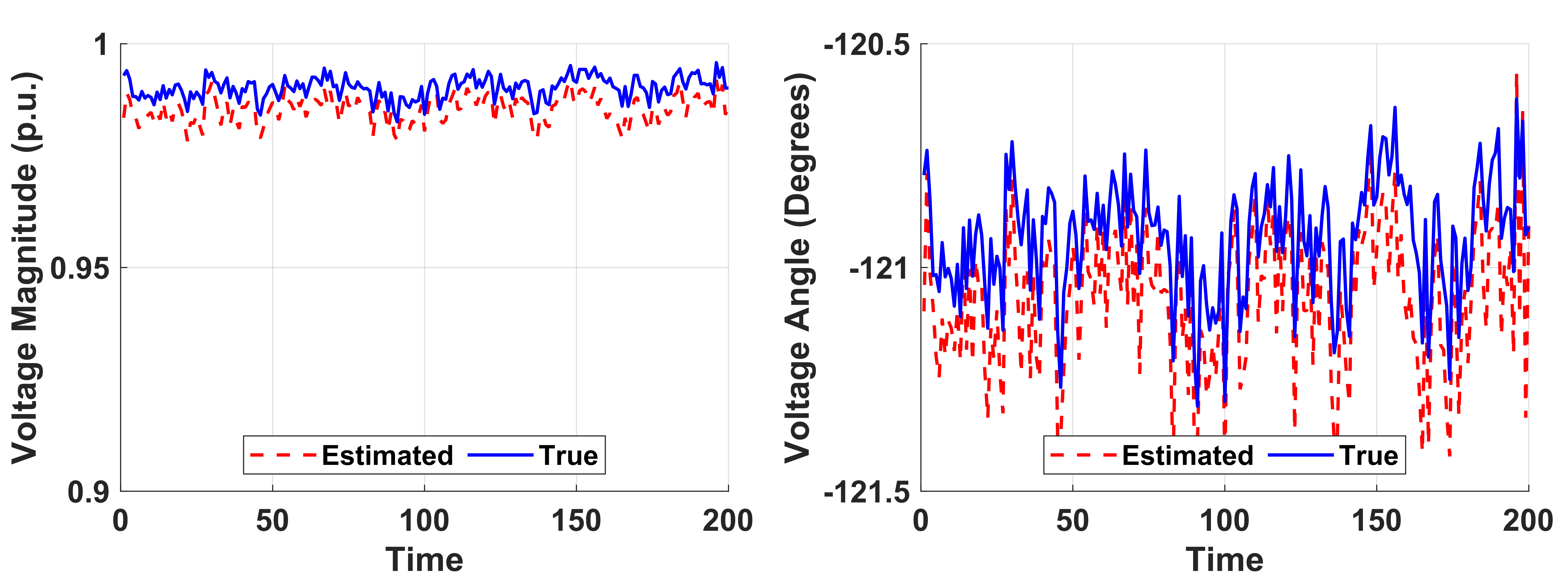}
      \caption{Voltage magnitude and angle at Bus 93.2}
      \label{fig:subplot4}
    \end{subfigure}
  \end{minipage}
  \caption{Results of Voltage magnitudes and angles at Bus11.1, Bus 54.2, Bus83.3, Bus93.2}
  \label{fig:estimating_all_buses}
\end{figure*}

\begin{figure*}[t]
  \begin{minipage}{0.32\textwidth}
    \centering
    \begin{subfigure}{\linewidth}
      \includegraphics[width=0.9\linewidth]{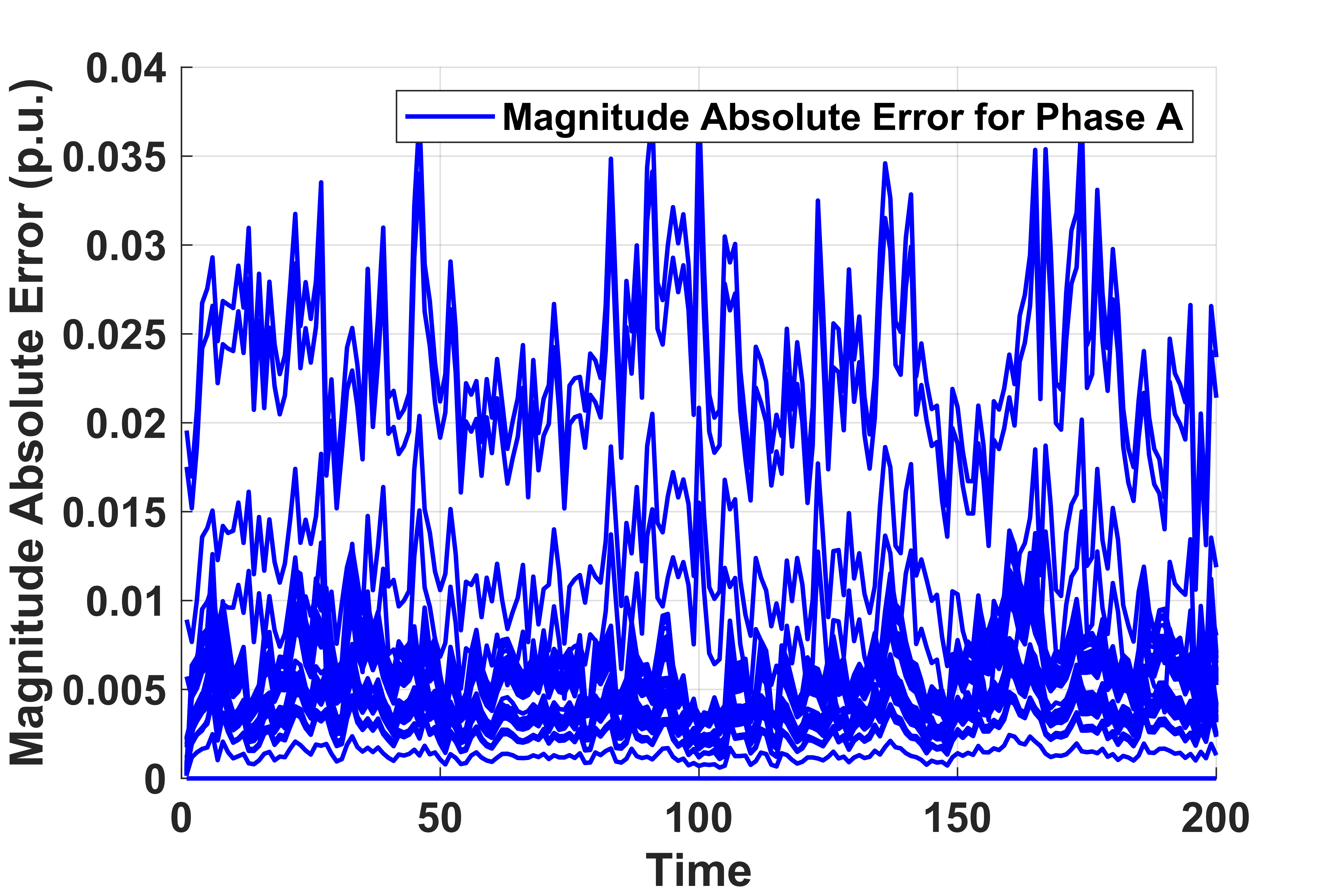}
      \label{fig:subplot1}
    \end{subfigure}
  \end{minipage}%
  \begin{minipage}{0.32\textwidth}
    \centering
    \begin{subfigure}{\linewidth}
      \includegraphics[width=0.9\linewidth]{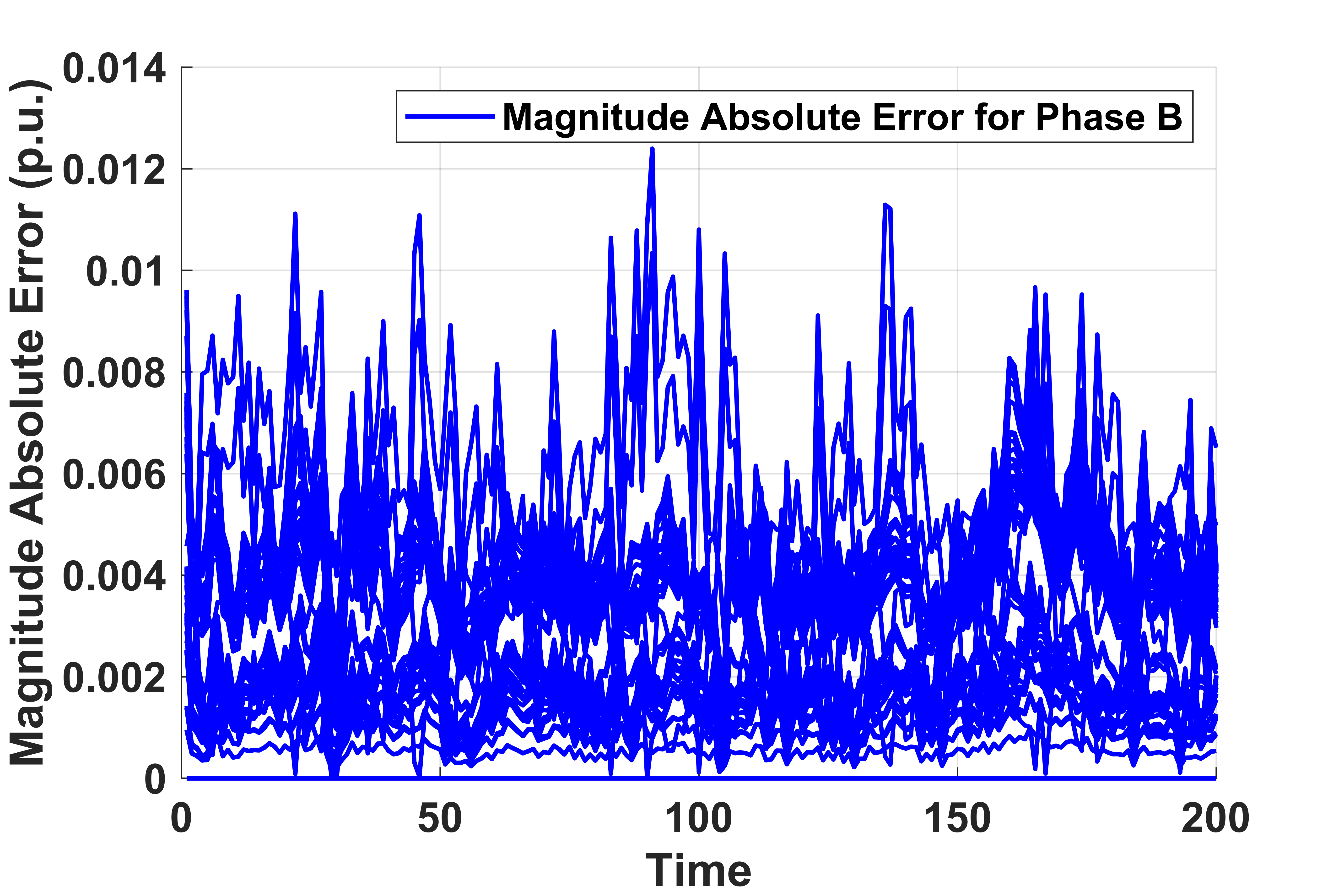}
      \label{fig:subplot2}
   \end{subfigure}
  \end{minipage}%
  \begin{minipage}{0.32\textwidth}
    \centering
   \begin{subfigure}{\linewidth}
      \includegraphics[width=0.9\linewidth]{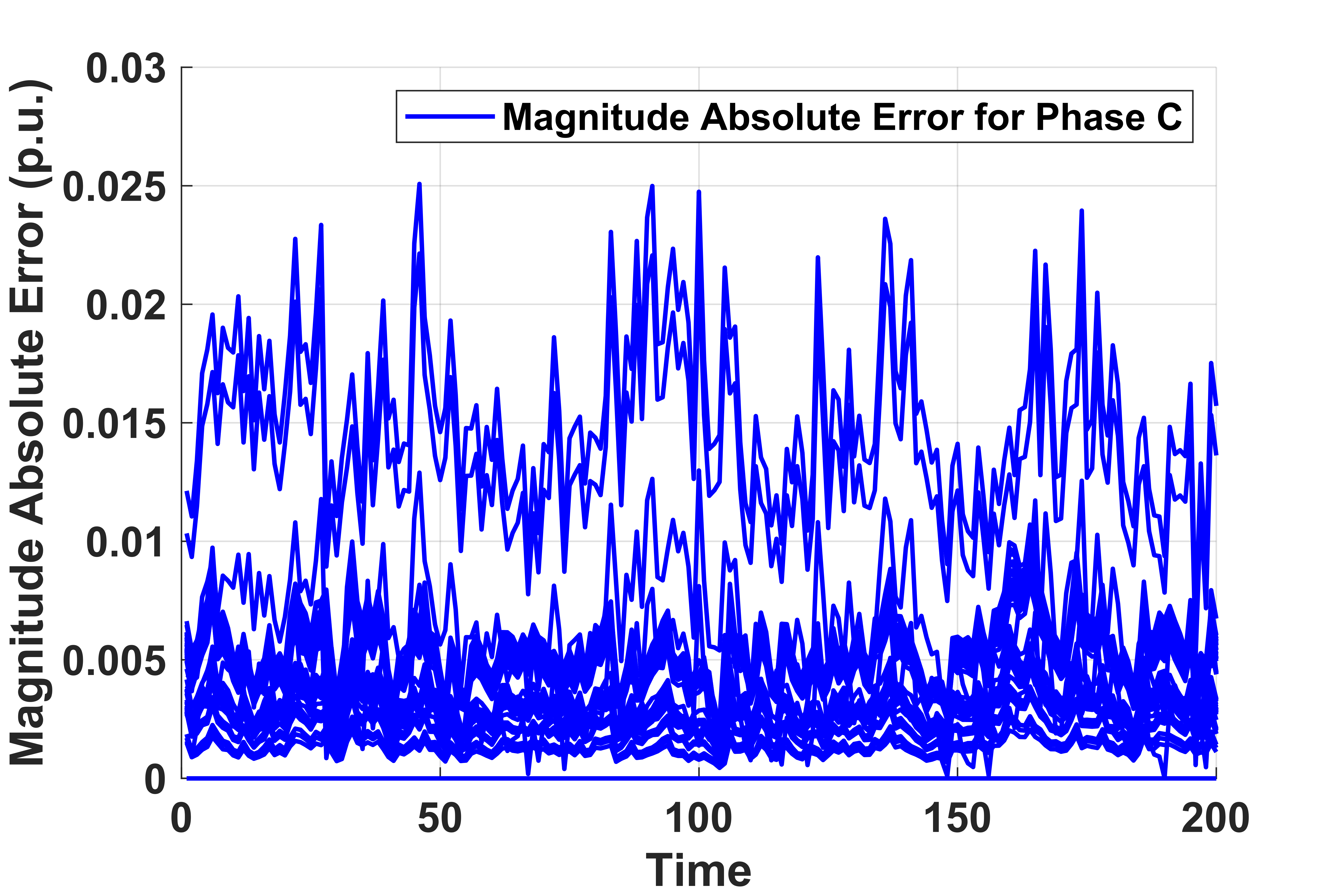}
      \label{fig:subplot3}
    \end{subfigure}
  \end{minipage}
  \caption{Absolute errors for voltage magnitudes for three phases}
  \label{fig:Mag_accuracy}
\end{figure*}

\begin{figure*}[!h]
  \begin{minipage}{0.32\textwidth}
    \centering
    \begin{subfigure}{\linewidth}
      \includegraphics[width=0.9\linewidth]{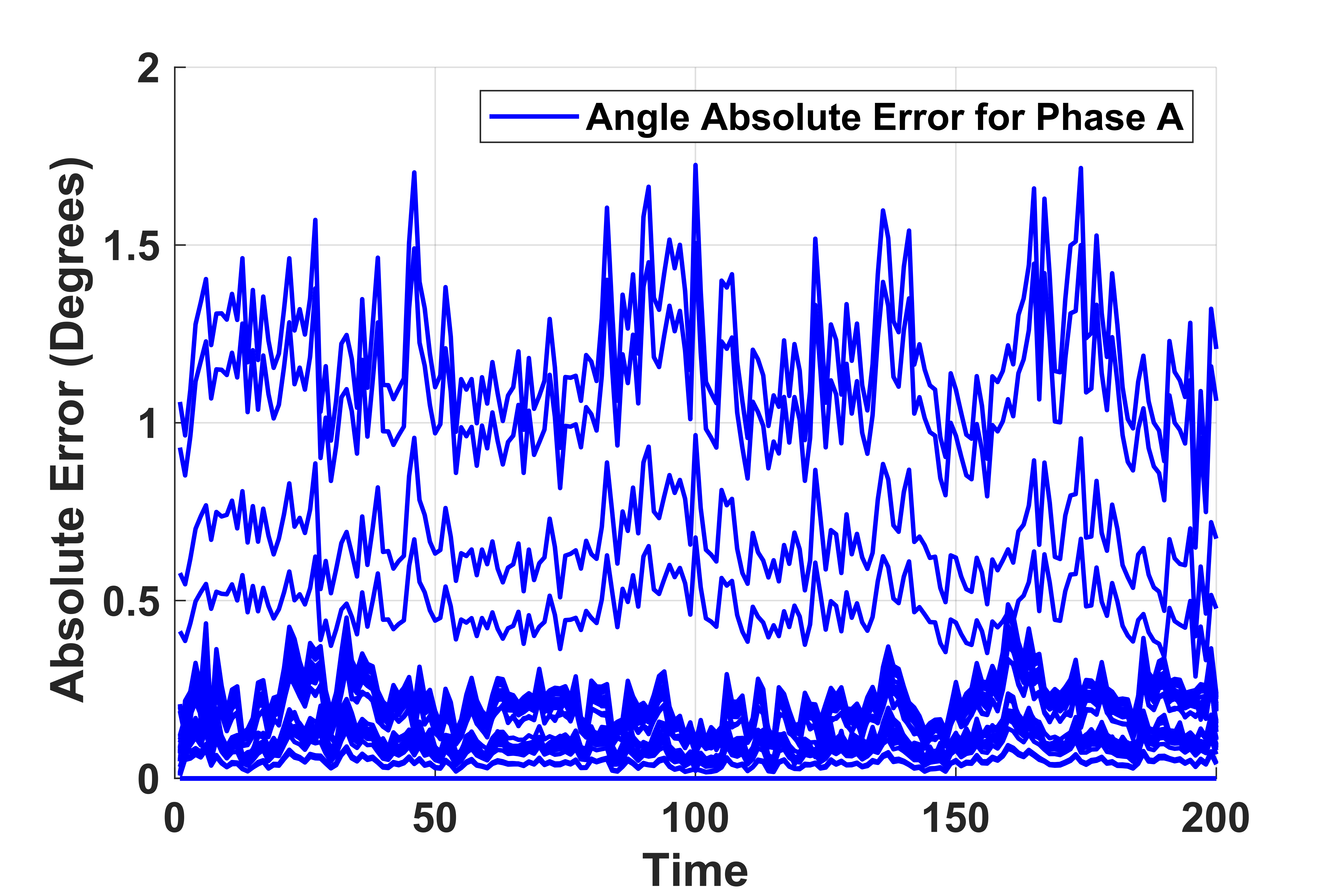 }
      \label{fig:subplot1}
    \end{subfigure}
  \end{minipage}%
  \begin{minipage}{0.32\textwidth}
    \centering
  \begin{subfigure}{\linewidth}
      \includegraphics[width=0.9\linewidth]{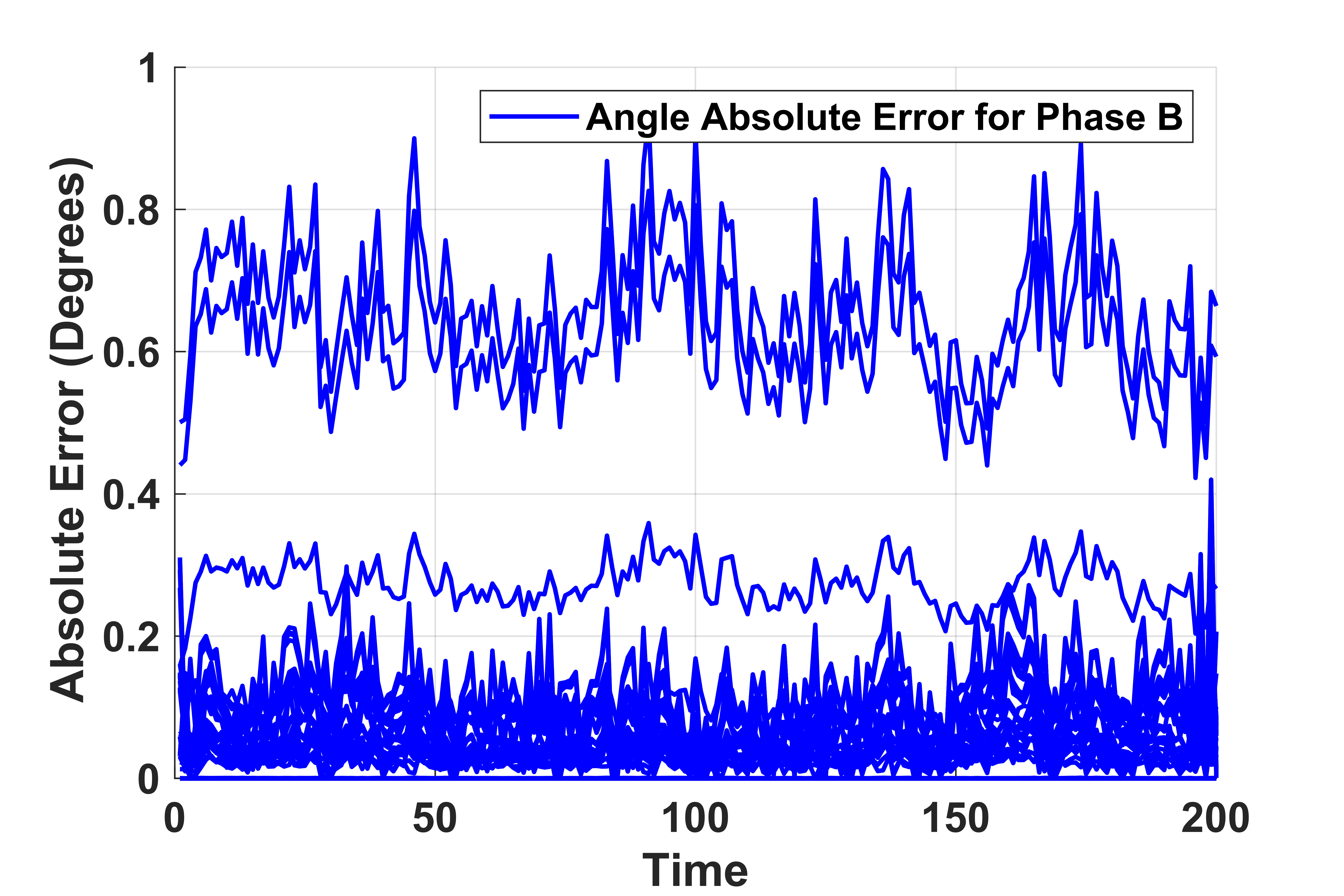 }
      \label{fig:subplot2}
   \end{subfigure}
  \end{minipage}%
  \begin{minipage}{0.32\textwidth}
    \centering
  \begin{subfigure}{\linewidth}
      \includegraphics[width=0.9\linewidth]{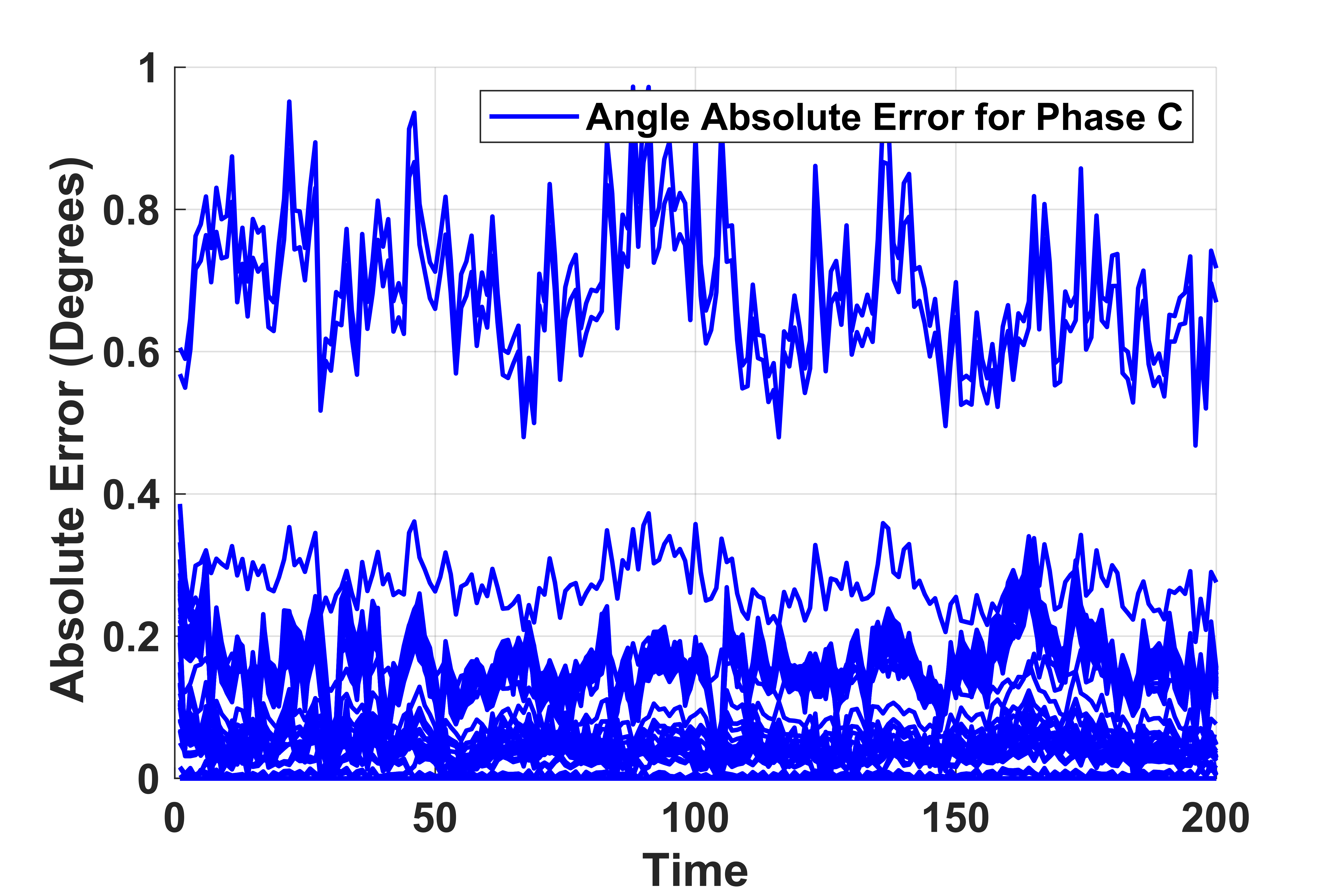 }
      \label{fig:subplot3}
    \end{subfigure}
  \end{minipage}
  \caption{Absolute errors for voltage angles for three phases}
  \label{fig:ang_accuracy}
\end{figure*}

 In the first case scenario, we used synthetic power profiles that mimicked the current of the UK power demands. The MAE of the power demand forecaster was around 0.006. Fig.~\ref{fig:2023} shows the MAE and RMSE for the three scenarios. The proposed structured approach shows a reasonable mean absolute error of less than 0.4 for the voltage magnitudes and less than 0.2 for the voltage angles across the three scenarios. As can be noticed, the voltage magnitudes and angles have high levels of irregularities that increase the difficulty level of estimating the states. The estimated voltage angles and magnitudes of selected buses are shown in Fig.~\ref{fig:estimating_all_buses}. These buses were selected since they have futuristic demand profiles. The accuracy of the estimated voltage magnitudes and angles for the 2023 scenario is shown in Fig.~\ref{fig:Mag_accuracy} \& Fig.~\ref{fig:ang_accuracy}.
 

 \subsection{\bfseries Scenario 2035}
The adoption of DERs and flexible loads at the UK LV network is predicted to increase. The increased deployment of rooftop solar panels in the 2035 UK power demands contributed to enhancing the power forecasting due to the reason of correlation between solar irradiance as an input feature and the generated powers at the LV level. The MAE of the power demand forecaster of this scenario was around 0.004. 


 \subsection{\bfseries Scenario 2050}
The visibility of the demand power forecasted in this scenario has been slightly enhanced since the adoption of rooftop solar panels is projected to increase. This enhancement took place since the solar irradiance was used as an input feature to the forecaster. The absolute error of the voltage magnitudes of the selected buses is less than 2\% for 200-time samples. It was noticed that the absolute error of the estimated states is steady among all three scenarios and normally distributed even with the huge differences in power consumption. 

\subsection{\bfseries Discussions}

There are various findings and key components in this systematic modeling of DSSE. This systematic modeling consists of various challenges that were addressed. We started with creating a time series of power demands that are suitable for deploying DSSE.

Synthetic power demands characteristics created an almost realistic dataset that mimics the future loading conditions of active distribution networks. Firstly, the power demands have correlations with weather information and price signals which can lead to have better understanding of consumption patterns. Also, this dataset includes key elements that are reshaping the future power demands of smart grids such as PV systems, BESS, HVAC systems, and EVs. Lastly, the dataset also has the consumption of individual smart meter units to help with further studies of the synchronization of the active distribution network measurements. 

This paper introduces an innovative power demand forecaster to overcome the limitations imposed by insufficient measurement devices. 
Since this structured approach introduced a comprehensive demand dataset, the forecaster takes advantage of the created demand dataset by incorporating the available limited real-time measurements of ADN, solar irradiance, and around 40\% of the available synchronized smart meter units. the load forecaster, a central component of this systematic modeling, demonstrates outstanding performance in handling time series data characterized by high irregularities. It outperforms state-of-the-art methods such as neural network transformers, LSTM, and CNN, boasting a Mean Absolute Error (MAE) of less than 0.06 and an R\textsuperscript{2} of approximately 84. 

Building on the foundation laid by the load forecaster and the dataset of ADN power demands, the systematic modeling progresses to DSSE based on FASE. Here, we used the dataset to create three load compositions representing present-day UK power demands, and projections for 2035 and 2050 to have a better assessment of the proposed strategy to enhance the performance of FASE. We found that through a thorough examination of the performance of FASE  based on Holt’s linear method, the smoothing parameters ($\alpha$ and $\beta$) could not be fixed for the whole time samples and for the unbalanced three phases. Thus, we came up with a strategy illustrated in equations~\eqref{eq:beta_update_phi} and~\eqref{eq:alpha_update_phi} that rely on the real-time measurement of the main branch currents. This has improved the accuracy of FASE. Considering multi-source measurements might improve the performance of this strategy~\cite{ref42}. The DSSE, coupled with the load forecaster and real-time measurements, consistently delivers promising results across all three scenarios (2023, 2035, 2050). Adding PMUs at optimal locations potentially increases the accuracy of the DSSE~\cite{ref43}
\vspace{-3pt}
\section{Conclusion}
This systematic modeling provides a valuable foundation for improving the accuracy and reliability of DSSE, particularly in the context of ADNs. We introduced synthetic power demand data with realistic characteristics, enabling a better understanding of consumption patterns and the impact of factors like weather and price signals, PV systems, BESS, HVAC systems, and EVs. The limitations imposed by the limited availability of measurement devices in the distribution network and the asynchronization challenge among smart meter units were addressed by an innovative power demand forecaster, WaveNet-LSTM, which demonstrated exceptional performance in handling the high level of irregularities in demand power consumption. We proposed a strategy for adaptive smoothing parameters in FASE based on Holt’s linear method that relies on real-time measurement of main branch currents to improve the estimation results. 

\bibliographystyle{IEEEtran}
\bibliography{refrencec.bib}

\end{document}